\documentclass[journal]{IEEEtran}
\usepackage{amsmath}
\usepackage{cite}
\usepackage{graphicx,amssymb}
\usepackage{amsmath,amsfonts,amssymb}

\usepackage[usenames]{color}
\usepackage{algorithm}
\usepackage{algorithmic}
\usepackage{epstopdf}
\usepackage{multirow}
\usepackage{stfloats}

\usepackage{comment}
\specialcomment{warning}{\color{red}}{\color{black}}
\specialcomment{new}{\color{blue}}{\color{black}}
\specialcomment{longcomment}{\color{gray}}{\color{black}}
\specialcomment{remove}{}{} \excludecomment{remove}

\newtheorem{proposition}{\textbf{Proposition}}

\begin{document}
 
\title{Wideband Collaborative Spectrum Sensing using Massive MIMO Decision Fusion}

\author{I.~Dey,~\IEEEmembership{Member,~IEEE}, D. Ciuonzo, \IEEEmembership{Senior Member,~IEEE}, and P.~Salvo~Rossi,~\IEEEmembership{Senior Member,~IEEE}
\thanks{Manuscript received 4th April 2019; revised 21th October 2019 and 30th January 2020; accepted 12th April 2020.}%
\thanks{I. Dey is with CONNECT, National University of Ireland, Maynooth (E-mail: indrakshi.dey@mu.ie).}
\thanks{D. Ciuonzo is with University of Naples ``Federico II'', Italy (E-mail: domenico.ciuonzo@unina.it).}
\thanks{P. Salvo Rossi is with Dept. of Electronic Systems, Norwegian University of Science and Technology (NTNU), Norway (E-mail: salvorossi@ieee.org).}
}

\markboth{IEEE Transactions on Wireless Communications,~Apr.~2020}{Dey \MakeLowercase{\textit{et al.}}:Wideband Collaborative Spectrum Sensing using Massive MIMO Decision Fusion}

\maketitle
\IEEEpeerreviewmaketitle

\begin{abstract}
In this paper, in order to tackle major challenges of spectrum exploration \& allocation in Cognitive Radio (CR) networks, we apply the general framework of Decision Fusion (DF) to wideband collaborative spectrum sensing based on Orthogonal Frequency Division Multiplexing (OFDM) reporting. 
At the transmitter side, we employ OFDM without Cyclic Prefix (CP) in order to improve overall bandwidth efficiency of the reporting phase in networks with high user density. On the other hand, at the receiver side (of the reporting channel) we device the Time-Reversal Widely Linear (TR-WL), Time-Reversal Maximal Ratio Combining (TR-MRC) and modified TR-MRC (TR-mMRC) rules for DF. 
The DF Center (DFC) is assumed to be equipped with a large antenna array, serving a number of unauthorized users competing for the spectrum, thereby resulting in a ``virtual'' massive Multiple-Input Multiple-Output (MIMO) channel. 
The effectiveness of the proposed TR-based rules in combating ($a$) inter-symbol and ($b$) inter-carrier interference over conventional (non-TR) counterparts is then examined, as a function of the Signal-to-Interference-plus-Noise Ratio (SINR). Closed-form performance, in terms of system false-alarm and detection probabilities, is derived for the formulated fusion rules. 
Finally, the impact of large-scale channel effects on the proposed fusion rules is also investigated, via Monte-Carlo simulations.
\end{abstract}


\IEEEpeerreviewmaketitle
\footnote{© 2020 IEEE.  Personal use of this material is permitted.  Permission from IEEE must be obtained for all other uses, in any current or future media, including reprinting/republishing this material for advertising or promotional purposes, creating new collective works, for resale or redistribution to servers or lists, or reuse of any copyrighted component of this work in other works.}
\section{Introduction}\label{S1}

\subsection{Motivation}\label{S1.1}

\IEEEPARstart{E}{merging} paradigms such as the Internet-of-Things (IoT) involve the coexistence of a multitude of communicating devices forming dense Radio Frequency (RF) communication networks.
These devices are generally expected to be small, low-powered, and in many relevant applications they will be in charge of transmitting sensed information to a centralized entity for further processing, so as to gather situation awareness of a certain phenomenon to be monitored.
This will lead to increased utilization of scarce resources, such as \emph{spectrum} and \emph{energy}.
As a result, dynamic spectrum management as well as energy efficient and environment-aware design should be jointly considered in the design \cite{yousefvand2015}.
Cognitive Radio (CR) promises a wonderland of available spectrum by accommodating more users in dense areas, by allowing co-existence of authorized Primary Users (PU) and unauthorized Secondary Users (SUs) on the same bandwidth \cite{haykin2005,devroye2006}, as recently demonstrated by its adoption (as an integral component) in IEEE 802.22 standard.
Recently, the (enhanced) concept of green CR has been introduced \cite{huang2015,yousefvand2017}, assuming all SUs possessing energy harvesting capabilities, thus able to provide self-sustainability and extend network lifetime \cite{ulukus2015}.

A major challenge in implementing (interweave) CR networks is the design of dynamic spectrum sensing and allocation algorithms for SUs without interfering the existing PU.
\emph{Spectrum sensing} accomplishes the task of dynamically inferring unused bandwidth portions, referred to as ``spectrum holes'' or  ``white spaces'' \cite{akyildiz2011,mitola1999}.

\subsection{Related Works to Spectrum Sensing}\label{S1.2}

Key spectrum sensing strategies can be grouped under \emph{four} categories based on the detection technique involved, i.e. ($a$) energy detection \cite{Urkowitz1967}, ($b$) coherent detection \cite{kay1998}, ($c$) cyclo-stationary feature detection \cite{Enserink1994} and ($d$) eigenvalue-based detection \cite{Zeng2009}.
These sensing strategies can be used to scan several frequency channels on an independent basis; a technique limited only to sense single (narrowband) channels \cite{romero2013}.
On the other hand, to enhance certain particularities (such as network throughput, sensing accuracy etc.), scanning multiple channels at the same time has long been considered \cite{romero2013,Hosseini2012}; a technique known as Wideband Spectrum Sensing (WSS).
Currently, energy detection is the most widely used approach both for narrowband sensing and WSS owing to its flexibility, robustness to implementation, and improvement in opportunistic throughput \cite{Urkowitz1967,quan2008}.
However, it is insufficient to detect presence of PU in a low Signal-to-Noise Ratio (SNR) region and within very short sensing time.\footnote{It is worth-mentioning that IEEE 802.22 requires CRs to sense PU signals as low as -114 dBm.}

WSS techniques aim at monitoring multiple bands jointly or sequentially \cite{Hosseini2012,quan2008icc}. 
Cooperative WSS schemes exploiting spatial diversity and improving sensing reliability have also been introduced in \cite{derakhtian2012,sun2016}.
However, the major showstopper for WSS is the high sampling rate required to sample the wide frequency range. 
To overcome this problem, researchers attempted the use of compressed sensing at sub-Nyquist rates \cite{tian2007,mishali2009,ariananda2012,sun2012,yen2013}.
However, the simplest WSS scheme proposed till date is Partial Band Nyquist Sampling (PBNS) \cite{sun2014}. 
It is based on the idea that a SU may not need information on spectral occupancy of all the frequency bands available; it will be interested in finding only one free band. 
A large number of WSS techniques are available in literature and their effectiveness have been tested against Additive White Gaussian Noise (AWGN) and narrow-band fading, but never against wideband channel effects like frequency-selective fading, interference between closely-spaced frequency bands and fast large scale channel effects.

An attractive solution to combat the detrimental effect of fading and shadowing is the centralized collaborative spectrum sensing \cite{unnikrishnan2008,quan2008}, where a Fusion Center (FC) collects the individual SU decisions and combines them to determine the presence/absence of the PU.
The Decision FC (DFC) implements array processing through multiple antennas (small, moderate and large array) even in case of single-antenna users \cite{Ciuonzo2012,salvorossi2013}.
Spectrum sensing from a joint spatio-temporal two-dimensional detection perspective is introduced in \cite{ding2013} using cognitive sensor networks, while decentralized alternatives have also been proposed in \cite{ding2012}. 
However, \emph{all these collaborative techniques only deal with narrowband sensing} and are yet to be explored for application to WSS.

\subsection{Related Works to Decision Fusion over Realistic Channels}\label{S1.3}

Distributed detection using Decision Fusion (DF) has been typically investigated in the context of Wireless Sensor Networks (WSNs).
Sub-optimum fusion rules have been applied to both Parallel Access Channel (PAC) \cite{Chen2006} and Multiple Access Channel (MAC) \cite{banavar2012} scenarios.
In case of a PAC architecture, the sensors are assigned orthogonal channels for reporting their decision whereas, in the case of MAC, sensors are allowed to transmit simultaneously.
The consequent interference observed in MAC scenarios is typically overcome with the exploitation of multiple antennas at the DFC.
Towards this end, decode-and-fuse and decode-then-fuse techniques were proposed and compared in \cite{Ciuonzo2012}.
The use of DF rules in the context of collaborative spectrum sensing has been introduced in \cite{peh2011} and \cite{umebayashi2012}.
Especially, \cite{peh2011} focuses on power allocation strategies in a scenario with one single SU transmitter and receiver pair cooperating to improve detection probability of PU activity. 

The effect of user cooperation and orthogonal transmission among SUs for spectrum sensing in CR scenarios, where the DFC is served with multiple antennas is introduced in \cite{salvorossi2013}.
It exploits array processing in order to improve performance through diversity gain from multiple antennas; the communication scenario in turn representing a virtual Multiple Input Multiple Output (MIMO) scenario.
Starting from the results in \cite{salvorossi2013} as a general framework for collaborative spectrum sensing (with the extension to multi-antenna SUs thoroughly investigated in \cite{patel2018robust}), in this paper, we consider a set of SUs transmitting over interfering reporting channels.
The appeal of the above setup has been recently experimentally confirmed through real measurement campaigns \cite{dey2019throughput} focusing on vitual MIMO DF set-up and \cite{dey2019tcom} concentrating on WSS based massive MIMO DF set-up.
We employ massive MIMO DF at the receiver side, following the success observed in \cite{ciuonzo2015,ding2018massive}.
Massive MIMO \cite{larsson2014} is a strong candidate for future communication networks, where the base station is equipped with a few hundred antenna elements. 
Its advantages include significant increase in capacity of multi-user networks, reduced latency and robustness to man-made breaches and intentional jamming.
Hence, we think that advantages offered by massive MIMO will be really useful in the context of collaborative WSS over multiple frequency bands.

\subsection{Related Works to Multi-carrier Techniques}\label{S1.4}

When employing WSS, if the frequency bands of interest are very closely spaced, the transmitted signal from the SUs will suffer from Inter-Symbol Interference (ISI) as well as Inter-Carrier Interference (ICI) during the reporting phase. This scenario will occur, especially if the available frequency bands belong to a RF communication system operating on a multi-carrier scheme like Orthogonal Frequency Division Multiplexing (OFDM), Generalized Frequency Division Multiplexing (GFDM), Filter Bank Multi-carrier (FBMC) modulation, which will potentially be integral parts of future communication systems \cite{guimaraes2013cooperative}. In this case, WSS will benefit from the fact that reliability of spectrum sensing is improved by utilizing the correlation of Cyclic Prefix (CP) in multi-carrier signals. However, a CP adds an extra overhead to the system. 

If DF is used for collaborative spectrum sensing over the closely spaced frequency bands belonging to an OFDM system, the system will suffer from additional complexity owing to high system knowledge requirement by the fusion rules. Removing CP reduces length of each packet and thereby improving overall bandwidth efficiency by a considerable amount and reducing the effect of channel aging due to network switching \cite{molisch2007iterative, beheshti2009equalisation}. Sadly, this can only be achieved at the expense of ISI as well ICI imposed by the channel transients. However, in the context of DF based WSS, performance can be improved even in presence of ISI and ICI, by formulating 
``large-MIMO''
version of each fusion rule. Such fusion rules can exploit linear increase in Signal-to-Interference-plus-Noise Ratio (SINR) with the number of receive antennas.

In case OFDM is combined with massive MIMO, addition of CP has detrimental effect on the available time for data transmission \cite{nsengiyumva2016}. 
This is because massive MIMO operates on Time Division Duplexing (TDD) mode.
The CP is a repetition of last samples in a symbol, which is appended at the beginning of the symbol to avoid ISI \cite{toeltsch2000}. But, this results in huge reduction of spectral efficiency almost canceling out the gain in capacity in multi-user networks. 
Recently academia has started looking into the possibility of eliminating (or shortening) CP lengths \cite{molisch2007iterative} at the cost of additional interference.
To mitigate this additional interference, several interference cancellation techniques have been proposed \cite{lim2006mimo, chen2009,ma2009two}.
However, if these cancellation methods are employed on a massive MIMO based communication system, the computation complexity increases exponentially with the increase in the number of receive antennas.

In absence of CP, TR-based techniques when applied to massive MIMO are found to improve performance both in wireless RF \cite{chen2016time} as well as underwater acoustic channels \cite{yang2012}. It is shown in \cite{aminjavaheri2017ofdm}, that saturation of SINR owing to residual ISI and ICI can be resolved using Time-reversed Maximal Ratio Combining (TR-MRC) and Time-reversed Zero Forcing (TR-ZF). Hence we think that using the benefits of TR on the DF side of collaborative WSS, we can ameliorate fusion performance even in presence of additional interference over multiple closely-spaced frequency bands. This is due to the fact that TR techniques can exploit information from the propagation environment to create a spatio-temporal resonance effect (or focusing effect) and to perform perfect deconvolution.

\subsection{Contribution and Paper Organization}\label{S1.5}
The main contributions of this manuscript are summarized as follows:
\begin{itemize}
\item {Starting from the results in \cite{salvorossi2013} (tackling the simpler narrowband case), we apply the general framework of distributed DF to the OFDM-based collaborative WSS to address the major challenges in wideband CR networks.
To the best of our knowledge, \emph{the application of large-array (massive) DFC is analyzed for the first time in the context of WSS}.
The aim is to exploit the asymptotic orthogonality of the interfering SU's local decisions observed from a DFC employing a massive array.
It is to be mentioned here that we rely on the assumption that number of antennas at the DFC is much larger than the number of transmitting SUs.}
\item {We employ OFDM on the transmitter side.
We \emph{eliminate} the use of CP in order to maintain \emph{high spectral efficiency} in dense network scenarios.
Additionally, we rely on the large-array gain of massive MIMO to average out the ISI and ICI introduced by the closely-spaced frequency bands in an OFDM-based system without CP.
In our opinion, the advantages offered by massive MIMO DF will be extremely beneficial in the context of WSS for multi-carrier based systems without CP. Advantages of large antenna array on the receiver side can be enjoyed without sacrificing spectral efficiency as well as facilitating low-complexity fusion rules and mitigating energy constraints.}
\item {We derive sub-optimum DF techniques with reduced complexity for the received signal at the DFC consisting of (i) Widely Linear (WL) rules, (ii) Standard Maximal Ratio Combining (MRC) and (iii) modified MRC (mMRC), generalizing to our setup those introduced in \cite{ciuonzo2015} for massive MIMO DF context. Additionally, (iv) Time-Reversal WL (TR-WL), (v) Time-Reversal MRC (TR-MRC), and (vi) modified TR-MRC (TR-mMRC) are designed with the intent of improving performance.
We highlight that TR-MRC has been introduced in \cite{han2016time} to mitigate ISI, ICI, Multiuser Interference (MUI) through spatio-temporal focusing, but has \emph{never been capitalized} for DF in WSNs or CRs.
Also, we remark that the reason for considering a set of (sub-optimal) rules originates from the need for  ``gracefully'' accommodating exponential complexity and high knowledge requirements (that is, ranging from naive (m)MRC to more sophisticated WL design principle, the latter taking into account reliability of the sensing process) of the optimum fusion rule (Sec.~\ref{S4.1}), both limiting its implementation.\footnote{Indeed, a high processing time at the DFC negatively impacts the overall latency required to reach a global decision on the spectrum availability and may thus reduce the overall efficiency of the CR network.}
``Large-MIMO" version of each fusion technique is developed such that they can truly exploit linear SINR increase with the array size.}
\item {Contrary to the studies in \cite{Ciuonzo2012, salvorossi2013, ciuonzo2015}, this is the first ever study of sub-optimum DF techniques against SINR regime, rather than the SNR regime. We include interference components along with noise for performance evaluation as the fusion rule statistics in many cases are proportional to channel coefficients and are dependent on the instantaneous Channel State Information (CSI)}. In addition to ISI and ICI, we investigate how large scale channel effects impact system performance of proposed fusion algorithms for collaborative WSS in CR networks.
\item {Closed-form expression, in terms of system false-alarm and detection probabilities, is derived when large array is employed at DFC, for the DF rules formulated here. We also examine potency of the TR techniques in combating ISI and ICI compared to other techniques from the context of WSS in OFDM-based systems operating without CP.}
\end{itemize}

This paper is organized as follows: Sec.~\ref{S2} introduces sensing and signal models. 
Sec.~\ref{S3} focuses on the considered channel models, along with the interference contributions due to both ISI and ICI effects. 
Sec.~\ref{S4} presents fusion techniques for collaborative WSS over massive OFDM-MIMO reporting channels. 
Sec.~\ref{S5} presents an extensive set of simulations for performance comparison under different scenarios. 
Finally, concluding remarks and further avenues of research are provided in Sec.~\ref{S6}.
\footnote{\emph{Notations:} Lower-case (resp. upper-case) bold letters denote vectors (resp. matrices), with $a_{k}$ (resp. $a_{n,m}$) representing the $k$th element (resp. $(n,m)$th element) of $\mathbf{a}$ (resp. $\mathbf{A}$); $(\cdot)^{t}$ denotes transpose and $\mathbb{E}\{\cdot\}$, $\mathbb{V}\{\cdot\}$, $\mathbb{R}\{\cdot\}$, $\angle(\cdot)$, $(\cdot)^\dagger$, and $||\cdot||$ represents mean, variance, real-part, phase, conjugate transpose and Frobenius norm operators, respectively; $\mathbf{I}_N$ denotes the $N \times N$ identity matrix; $\mathbf{0}_N$ (resp. $\mathbf{1}_N$) denotes the null (resp. ones) vector of length $N$; $\underline{\mathbf{a}}$ (resp. $\underline{\mathbf{A}}$) denotes the augmented vector (resp. matrix) of $\mathbf{a}$ (resp. $\mathbf{A}$) i.e., $\underline{\mathbf{a}} \triangleq [\mathbf{a}^t~~\mathbf{a}^{\dagger}]^t$ (resp. $\underline{\mathbf{A}} \triangleq [\mathbf{A}^t~~\mathbf{A}^{\dagger}]^t$); $P(\cdot)$ and $p(\cdot)$ are used to denote probability mass functions (PMF) and probability density functions (PDF); $\mathcal{N}({\mu},{\Sigma})$ and $\mathcal{N}_{\mathbb{C}}({\mu},{\Sigma})$ denote normal distribution and circular symmetric complex normal distribution with mean vector ${\mu}$ and covariance matrix ${\Sigma}$ respectively; ${Q}(\cdot)$ is used to denote the complementary cumulative distribution function (CCDF) of standard normal distribution; $\chi_{k}^{2}$ (resp. $\chi_{k}^{'2}(\xi)$) denotes a chi-square (resp. a non-central chi-square) distribution with $k$ degrees of freedom (resp. and non-centrality parameter $\xi$) and `$\text{mod}~L$' refers to the modulo-operation which returns the remainder after division by $L$.} 
\section{System Model} \label{S2}

\begin{figure}[t]
\begin{center}
 \includegraphics[width=0.9\linewidth]{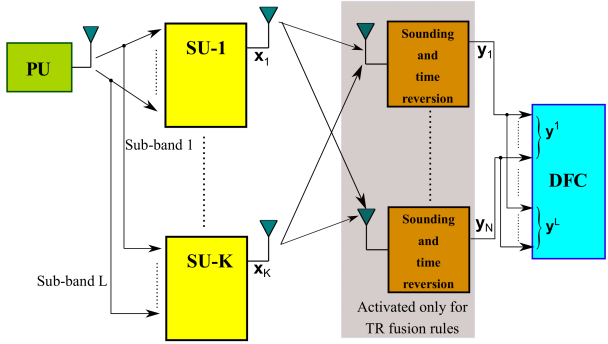}
\end{center}
\vspace*{-2mm}
\caption{Collaborative WSS through massive MIMO DF for the $l$th sub-carrier.}
\label{FIG30}
\vspace*{-3mm}
\end{figure}

In this paper, we consider an OFDM-based cognitive system with one PU and $K$ unauthorized SUs that want to transmit in the licensed spectrum, here divided into $L$ frequency bands, provided that the authorized PU is silent.
Although there are several definitions of a vacant frequency band (viz. white space), we can generally postulate that a frequency band is unoccupied (occupied) if the filtered radio signal within this band is composed of only noise (signal plus noise).
The spectrum sensing scenario considered herein for the $l$th sub-carrier is illustrated in Fig.~\ref{FIG30}, where SU-$k$ represents $k$th SU.
In what follows, we focus separately on the sensing model of each SU (Sec.~\ref{S2.1}) and the (received) signal model (Sec.~\ref{S2.2}), concerning the reporting phase to the DFC.

\subsection{Sensing and Local Decision Model} \label{S2.1}
The $k$th SU ($k \in \mathcal{K} \triangleq \{1, 2, \dotso, K\}$), equipped with a single antenna, senses those $L$ frequency bands (viz. the whole spectrum) and takes a local (1-bit) decision corresponding to $l$th frequency PU state (being silent or active).
The local decision on $l$th frequency band is then mapped to a Binary Phase-Shift Keying (BPSK) modulated symbol, $x_k^l \in \mathcal{X} \triangleq \{+1, -1\}$ transmitted by the $k$th SU on the $l$th sub-carrier. Therefore, each SU transmits a total of $L$ bits in each OFDM symbol.

Let the hypothesis that PU is silent (resp. active) on the $l$th sub-carrier be denoted by $\mathcal{H}_0^l$ (resp. $\mathcal{H}_1^l$).
We assume that the local sensing and decision process at the $k$th SU over the $l$th sub-carrier is fully described by the conditional probabilities $P(x_k^l|\mathcal{H}_i^l)$.
Specifically, we denote the probability of detection and false-alarm at $k$th SU for $l$th sub-carrier as $P_{D, k}^l \triangleq P(x_k^l = 1|\mathcal{H}_1^l)$ and $P_{F, k}^l \triangleq P(x_k^l = 1|\mathcal{H}_0^l)$, respectively.
Finally, for compactness, let $\mathbf{x}_{k}\triangleq\begin{bmatrix}x_{k}^{1} & \cdots & x_{k}^{L}\end{bmatrix}^{t}$ (resp. $\mathbf{x}^{l}\triangleq\begin{bmatrix}x_{1}^{l} & \cdots & x_{K}^{l}\end{bmatrix}^{t}$) be the set of local decisions transmitted from $k$th SU on the $L$ sub-carriers (resp. from all the $K$ SUs on $l$th sub-carrier).

\subsection{Signal Model} \label{S2.2}

We assume that $K$ SUs transmit their $L$-dimensional decision vectors independently on the reporting channel.
This assumption is critical to our work as \emph{we do not consider user cooperation among the SUs} for transmitting their decisions on the reporting channel, as is done in \cite{salvorossi2013} to increase spatial diversity.
Still, the chosen setup appealingly allows to obtain a latency in the collection of SUs decisions at the DFC which \emph{does not grow} with $K$.

\begin{figure}[b]
\begin{center}
 \includegraphics[width=0.9\linewidth]{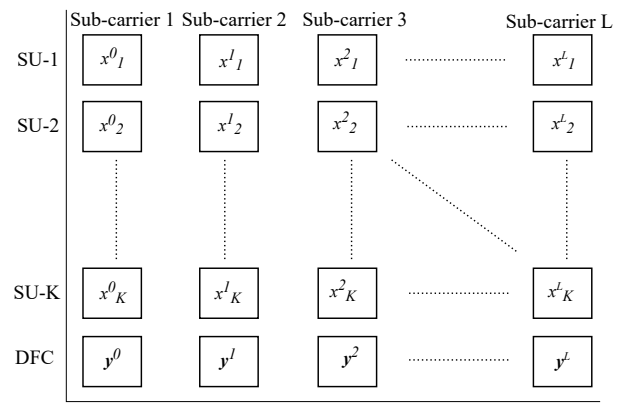}
\end{center}
\vspace*{-2mm}
\caption{Structure of transmission frame over $L$ sub-carriers.}
\label{FIG40}
\vspace*{-3mm}
\end{figure}

The DFC is equipped with $N$ receive antennas over a wireless flat-fading MAC in order to exploit diversity and combat signal attenuation due to small scale fading and large scale shadowing; this set-up determines a distributed or `virtual' massive MIMO channel.
The generic $N$-length received vector at the DFC is denoted by $\mathbf{y}^l \triangleq \big(y_1^l, y_2^l, \dotso, y_N^l\big)^t$ where $y_n^l$ is the signal received by the $n$th receive antenna on the $l$th sub-carrier.
The transmission frame for the general case of $K$ SUs over $L$ sub-carriers is shown in Fig.~\ref{FIG40}.
A \emph{large-array} configuration is considered here, that is $N >> K$; however, the formulation and results are applicable to any MIMO-DF framework. In summary, the communication process on the reporting channel for the $l$th sub-carrier may be viewed as a $K \times N$ massive MIMO system.

In this paper, we consider a CR-like scenario where massive MIMO DF based collaborative WSS is employed to closely spaced frequency bands belonging to a multi-carrier system like an OFDM-based system.
In a typical OFDM-based system, coherence time of the channel can be divided into the training period and data transmission period. In this paper, we concentrate on the data transmission period only and consider eliminating the CP altogether from this period, as is done in \cite{chen2016time}. 
In that case, the channel between the SUs and the DFC will suffer from ISI and ICI. 
Therefore, we include both additive noise and interference in our signal model and mathematically analyze SINR performance of sub-optimum DF rules in absence of CP in an OFDM signal.

We also assume perfect synchronization at the DFC.
Assuming perfect timing and frequency synchronization, the discrete-time signal model (after matched filtering and sampling) for the received signal at the DFC is given by,
\begin{align} \label{eq1}
\mathbf{y}^l = \sqrt{\rho^l} \mathbf{G}^l \mathbf{x}^l + \mathbf{w}^l + \mathbf{\Psi}^l
\end{align}
where $\mathbf{y}^l \in {\mathbb{C}}^N$, $\mathbf{G}^l \in {\mathbb{C}}^{N \times K}$, $\mathbf{x}^l \in {\mathbf{\chi}}^K$, $\mathbf{w}^l \sim \mathcal{N}_{\mathbb{C}} (\mathbf{0}_N, \sigma_{w, l}^2 \mathbf{I}_N)$ and $\mathbf{\Psi}^l \sim \mathcal{N}_{\mathbb{C}} (\mathbf{0}_N, \psi_{l}^2 \mathbf{I}_N)$ are the received signal vector, the channel matrix, the transmitted signal vector, the noise vector and the interfering signal vector respectively. In (\ref{eq1}), the constant $\rho^l$ denotes the energy spent by a generic SU during the reporting phase. The component for interference $\mathbf{\Psi}^l$ in (\ref{eq1}) arises from the combination of ISI among symbols carrying decisions of all $K$ SUs on each sub-carrier, and ICI due to nearby sub-carriers. The matrix $\mathbf{G}^l$ includes all the samples of the Channel Impulse Response (CIR) between the users and the DFC on the $l$th sub-carrier.

The DFC at receiver side is in charge of providing a reliable decision about the activity of the PU (i.e. $\mathcal{H}^1,\ldots,\mathcal{H}^L$) on the basis of the superimposed received (via the wireless channel) decisions taken locally by the SUs independently on each sub-carrier (i.e. $\mathbf{y}^1,\ldots,\mathbf{y}^L$).
In this way, the system finally gets a picture of the white spaces available, which are then exploited by a CR coordinator. 

\footnotetext[1]{Constance of ${\lambda_k^l}$ over $n$ is justified since the SU-DFC distance is typically much higher with respect to the inter-antenna distance.} 
\section{Channel and Interference Modeling} \label{S3}

\subsection{Channel Model} \label{S3.1}

The generic channel coefficient vector $\mathbf{g}^l_{n, k}$ is expressed as, $\mathbf{g}^l_{n, k} = \sqrt{\lambda_k} \mathbf{h}^l_{n, k}$ for $(n = 1, 2, \dotso N, l = 1, 2, \dotso L)$, where ${\lambda_k^l}$\footnotemark[1] models the geometric attenuation and shadow-fading and remains constant over $n$ and $l$. Each of the fast fading CIRs, $\mathbf{h}^l_{n, k}$ can in terms be modeled as linear time-invariant Finite Impulse Response (FIR) filters with the order of $Z$, i.e., $\mathbf{h}^l_{n, k} = \big({h}^l_{n, k}(0), \dotso, {h}^l_{n, k}(Z - 1)\big)^t$ and $\mathbf{h}^l_{n, k} \sim \mathcal{N}_{\mathbb{C}} (0, \text{diag}(\mathcal{B}^l_k))$ where the vector $\mathcal{B}^l_k = \big(\beta^l_k(0), \dotso, \beta^l_k(Z - 1)\big)^t$ is the power delay profile (PDP) of the channel model. Throughout this paper, we assume a normalized PDP i.e. $\sum_{z = 0}^{Z - 1} \beta^l_k (z) = 1$. Based on these assumptions we have, $\mathbf{G}^l = \mathbf{H}^l \sqrt{\mathbf{D}}~(l = 1, 2, \dotso, L)$ where, $\mathbf{G}^l \in {\mathbb{C}}^{N \times K}$ denotes the matrix of the generic channel coefficients with $\mathbf{g}^l_{n, k}$ as the vector element on the $n$th row and the $k$th column, $\mathbf{H}^l \in {\mathbb{C}}^{N \times K}$ denotes the matrix of the fast-fading coefficients with $\mathbf{h}^l_{n, k}$ as the vector element on the $n$th row and the $k$th column and $\mathbf{D} \in {\mathbb{C}}^{K \times K}$ is a diagonal matrix where $d_{k, k} = \lambda_k$.

Throughout this paper, we consider that the DFC has perfect knowledge of the CSI.
Accordingly, we assume that part of the coherence interval is used for training to perfectly estimate the channel and to establish the carrier frequency and timing synchronization.
For example, if $\tau^l$ and $\tau_{\text{sync}}$ be the number of pilot and synchronization symbols and $(\tau_c^l - \tau^l - \tau_{\text{sync}})$  symbols are used for DF task over each sub-carrier, then $\tau_c^l$ is the total number of symbols transmitted within the channel coherence interval over the $l$th sub-carrier.
During the training phase, all SUs transmit mutually orthogonal pilot sequences of length $\tau^l$ over each sub-carrier.

It is to be noted here that we consider only a single-slot (i.e. $\tau_c^l=1$) reporting phase in our work and therefore, we do not consider the impact of multiple information symbols (providing ``time-diversity") on the sensing method, but leave this interesting generalization for future work.

\subsection{Favorable Propagation}\label{S3.2}

If we denote the $k$th column of the channel matrix $\mathbf{G}^l$ as $\mathbf{g}_k^l$, where the vectors $\mathbf{g}_k^l$, $k \in \mathcal{K}$, are mutually independent complex-valued Gaussians with moments, $\mathbb{E}\{\mathbf{g}^l_{k}\} = \mathbf{0}_N$ and $\mathbb{E}\{\mathbf{g}^l_{k}(\mathbf{g}^l_{k})^{\dagger}\} = \lambda_k^l*[\text{diag}(\mathcal{B}^l_k)]*\mathbf{I}_N$. Thus the so-called favorable propagation conditions \cite{larsson2014} hold, i.e., $\frac{1}{N} (\mathbf{G}^l)^{\dagger} \mathbf{G}^l \approx \mathbf{D}_g^l$ for $N>>K$, where $\mathbf{D}_g^l = \mathbf{D}*[\text{diag}(\mathcal{B}^l_k)] = [\text{diag}(\mathbf{\Lambda})]*[\text{diag}(\mathcal{B}^l_k)]$, $\mathcal{B}^l_k = \big(\beta^l_k(0), \dotso, \beta^l_k(Z - 1)\big)^t$ and $\mathbf{\Lambda} = (\lambda_1, \dotso, \lambda_k)$.

\subsection{Modified System Model}\label{S3.3}

Here we develop the time-reversed (TR) version of the channel model in order to formulate the TR-based fusion rules. Let us denote the TR variant of the channel matrix, denoted by $\breve{\mathbf{G}}^l$ on the $l$th sub-carrier. Each element of $\breve{\mathbf{G}}^l$ in this case can be expressed as, $\breve{\mathbf{g}}^l_{n,k} = \sqrt{\lambda_k}\breve{\mathbf{h}}^l_{n,k}$ for $(n = 1, \dotso, N,~l = 1, \dotso, L)$, where $\breve{\mathbf{h}}^l_{n,k}$ is the TR version of ${\mathbf{h}}^l_{n,k}$, given by, $\breve{\mathbf{h}}^l_{n,k} = (h^l_{n,k}(Z - 1), \dotso, h^l_{n,k}(0))^t$. Essentially, $\breve{\mathbf{h}}^l_{n,k}$ becomes $\breve{\mathbf{h}}^l_{n,k} \sim \mathcal{N}_{\mathbb{C}}(0, \text{diag}(\breve{\mathcal{B}}^l_k))$, where $\breve{\mathcal{B}}^l_k = [\beta_k^l (Z -1), \beta_k^l (Z -2), \dotso, \beta_k^l (0)]$ is the TR version of the channel PDP and $\sum_{z = 0}^{Z - 1} \beta_k^l (z) = 1$. Based on these assumptions, we have $\breve{\mathbf{G}}^l = \breve{\mathbf{H}}^l \sqrt{\mathbf{D}}$ where $\breve{\mathbf{H}}^l$ denotes the TR channel matrix containing the fading coefficients and definition of $\mathbf{D}$ remains same as in Subsection~\ref{S3.1}.

Since, we are assuming favorable propagation condition, the channel matrices ${\mathbf{G}}^l$ are pairwisely orthogonal. Hence their time-reversed versions are also pairwisely orthogonal to each other and therefore, we can write, $\frac{1}{N}(\breve{\mathbf{G}}^l)^{\dagger} \breve{\mathbf{G}}^l \approx \breve{\mathbf{A}}^l$. In this case, $\breve{\mathbf{A}}^l = \mathbf{D}*[\text{diag}(\breve{\mathcal{B}}^l_k)]$. At the same time, the channel matrix will also be pairwisely orthogonal to its time-reversed version. Hence, as $N \to \infty$, $(\breve{\mathbf{G}}^l)^{\dagger} {\mathbf{G}}^l \approx \frac{1}{N} \mathbf{F}^l$. In this case, ${\mathbf{F}}^l = \Big(\sqrt{\mathbf{D}}*\Big[\text{diag}(\sqrt{\breve{\mathcal{B}}^l_k})\Big]\Big)^{\dagger} * \Big(\sqrt{\mathbf{D}}*\Big[\text{diag}(\sqrt{{\mathcal{B}}^l_k})\Big]\Big)$.
\subsection{Inter-Carrier Interference (ICI)}\label{S3.4}

Here, we derive the ICI power from a given sub-carrier $q$ on sub-carrier $p$, after reception at the DFC. Defining the sub-carrier distance as $d_{pq} \triangleq \{(q - p)~\text{mod}~L\}$, asymptotic ICI power from sub-carrier $q$ with the modulo-$L$ distance $d_{pq}$ from sub-carrier $p$, as the number of DFC antennas tend to infinity ($N \to \infty$) can be obtained as, $\lim_{N \to \infty} \big\{\psi^2_{p, q, \text{ICI}}\big\} = \sum_{k = 1}^K \big[\mathbb{E}\big\{\mathbf{h}_{n, k}^p * \big[\mathbf{H}_{n, k}^{\text{ICI}}\big]^{pq}\big\}\big]^2$, such that,
\begin{align} \label{eq7a}
\mathbf{H}_{n, k}^{\text{ICI}} = 
\begin{pmatrix}
  h_{n,k}(0) & 0 & \cdots & 0 \\
  h_{n,k}(1) & h_{n,k}(0) & \cdots & 0 \\
  \vdots  & \vdots  & \ddots & \vdots  \\
  h_{n,k}(Z-1) & h_{n,k}(Z-2) & \cdots & 0 \\
  0 & h_{n,k}(Z-1) & \cdots & 0 \\
  \vdots  & \vdots  & \ddots & \vdots  \\
  0 & 0 & \cdots & h_{n,k}(0)
 \end{pmatrix}.
\end{align}
where $\mathbf{H}_{n, k}^{\text{ICI}}$ is a lower triangular Toeplitz matrix that collects all the channel coefficients contributing ICI between $n$th DFC antenna and $k$th SU and is independent of $l$. Therefore,
\begin{align} \label{eq7}
\lim_{N \to \infty} \big\{\psi^2_{p, q, \text{ICI}}\big\} 
=& \sum_K \Bigg|\mathbb{E}\Bigg\{\sum_{z = 0}^{Z - 1}\sum_{z' = 0}^{Z - 1} h_{n, k}^{*} (z') h_{n, k}(z) \nonumber\\
&~~~~~~~~~~\times\,e^{- j2\pi(zq - q - z'p + p)} w(z)\Bigg\}\Bigg|^2 \nonumber\\
=& \sum_K \Bigg|\mathbb{E}\Bigg\{\sum_{z = 1}^{Z} \mathbf{\beta}_k^{p} (z)\,e^{- j2\pi z d_{pq}}\Bigg\}\Bigg|^2 \nonumber\\ 
=& \sum_K \big| \overline{\mathbf{\beta}}_k^{p} (d_{pq})\big|^2
\end{align}
where $\overline{\mathbf{\beta}}_k^{p} (d_{pq})$ contains all the Discrete Fourier Transform (DFT) samples of the channel PDP between the $p$th and the $q$th sub-carriers and $w(z)$ is the unity window function. Therefore, for all the $L$ sub-carriers contributing to ICI can be given by,
\begin{align} \label{eq8}
\lim_{N \to \infty} \big\{\psi^2_{l, \text{ICI}}\big\} &= \sum_K \big[\big| \overline{\mathbf{\beta}}^{p}_k (d_{1l})\big|^2 + \dotso + \big| \overline{\mathbf{\beta}}^{p}_k (d_{Ll})\big|^2\big] \nonumber\\
&= \sum_{k = 1}^K \sum_{p = 1}^{L} \big| \overline{\mathbf{\beta}}^{p}_k (d_{pl})\big|^2 ~~~\text{for}~~p \neq l
\end{align}
since, we are neglecting the presence of any self-interference on the sub-carrier of interest. It is to be clarified here that ICI depends on the distance between the subcarriers and the ICI power experienced on the $l$th subcarrier is given by, $\psi^2_{l, \text{ICI}}$, which is dependent on $l$.

\subsection{Inter-Symbol Interference (ISI)}\label{S3.5}

Next, we derive the ISI power within the $l$th sub-carrier after reception at the DFC. Advancing in the same way as in case of ICI, as $N \to \infty$, the ISI power can be calculated as, $\lim_{N \to \infty} \big\{\psi^2_{l, \text{ISI}}\big\} = \sum_{k = 1}^K\big[\mathbb{E}\big\{\mathbf{h}_{n, k}^l * \big[\mathbf{H}_{n, k}^{\text{ISI}}\big]^{lq}\big\}\big]^2$, where $\mathbf{H}_{n, k}^{\text{ISI}}$ is the $L \times L$ frequency domain ISI matrix between the $n$th DFC antenna and the $k$th SU. It is an upper triangular Toeplitz matrix in nature and can be expressed as,
\begin{align} \label{eq10a}
\mathbf{H}_{n, k}^{\text{ISI}} = 
\begin{pmatrix}
  0 & \cdots & h_{n,k}(Z-1) & \cdots & h_{n,k}(1) \\
  0 & \cdots & 0 & \cdots & h_{n,k}(2) \\
  \vdots & \ddots & \vdots & \ddots & \vdots  \\
  0 & \cdots & \cdots & \cdots & h_{n,k}(Z-1) \\
  0 & \cdots & \cdots & \cdots & 0 \\
  \vdots & \ddots & \vdots & \ddots & \vdots  \\
  0 & \cdots & \cdots & \cdots & 0
 \end{pmatrix}.
\end{align}
Therefore,
\begin{align} \label{eq10}
\lim_{N \to \infty} \big\{\psi^2_{l, \text{ISI}}\big\} =& \sum_K \Bigg|\mathbb{E}\Bigg\{\sum_{z = 0}^{Z - 1}\sum_{z' = 0}^{Z - 1} h_{n, k}^{*} (z') h_{n, k}(z)  \nonumber\\
&~~~~~~~~~~\times \,e^{- j2\pi(z - z')p} w(z)\Bigg\}\Bigg|^2 \nonumber\\
=& \sum_K \Bigg|\sum_{z = 0}^{Z - 1} z \mathbf{\beta}^{l}_k (z)\Bigg|^2 = \sum_{k = 1}^K {\overline{\mathbf{\tau}}}_{l, k}^2
\end{align}
where $\overline{\mathbf{\tau}}_{l, k} = \sum_{z = 0}^{Z - 1} z \mathbf{\beta}^{l}_k (z)$ is the average delay spread on the $l$th sub-carrier channel between the $k$th SU and the DFC.

It is noteworthy that ICI and ISI will have impact on both the sensing and reporting phase. Errors in sensing due to interference will also leak into the reporting phase. But in this paper, since we are concerning ourselves with the reporting phase, we assume no errors in sensing before being transmitted on the reporting channel. Evaluating effect of ISI and ICI on the sensing phase will be considered in our future work.
\vspace{-2mm}
\section{Wideband Collaborative Spectrum Sensing} \label{S4}

Collaborative sensing is the process of making a final decision on the white space available for the network based on the sensing data, $\mathbf{y}^l$, that is collected from $K$ SUs. Here, we consider different fusion rules to be employed at the virtual massive MIMO DFC discussed in the next sub-sections. It is worth-mentioning here that the observation and reporting on each of the unused frequency bands is correlated with that on any of the other frequency bands. However, for the sake of simplicity, we assume spectrum sensing and reporting on each frequency band to be independent and decoupled for the formulations done in this paper. 

\subsection{Optimum Rule}\label{S4.1}

The test statistics for energy detector for each $l$th sub-channel is computed as,
\begin{align} \label{eq4}
\mathbf{\Gamma}_{\text{opt}}^l \triangleq \text{ln} \Bigg[\frac{p\big(\mathbf{y}^l | \mathbf{G}^l, \mathcal{H}_1^l \big)}{ p\big(\mathbf{y}^l | \mathbf{G}^l, \mathcal{H}_0^l \big)}\Bigg]~~\substack{\hat{\mathcal{H}} = \mathcal{H}_1 \\ > \\ < \\ \hat{\mathcal{H}} = \mathcal{H}_0}~~\gamma^l  
\end{align}
where $\hat{\mathcal{H}}$, $\mathbf{\Gamma}_{\text{opt}}^l$ and $\gamma^l$ denote the hypotheses, the log-likelihood ratio (LLR) and the threshold with which the LLR is compared to. Exploiting the independence of $\mathbf{y}^l$ from $\mathcal{H}_i^l$, given $\mathbf{x}^l$, an explicit expression of the LLR in (\ref{eq4}) is obtained as,
\begin{align} \label{eq5}
&\mathbf{\Gamma}_{\text{opt}}^l = \ln \Bigg[\frac{\sum_{\mathbf{x}^l} \exp\Big(-\frac{||\mathbf{y}^l - \sqrt{\rho^l}\mathbf{G}^l \mathbf{x}^l||^2}{\sigma^2_{e, l}}\Big) P\big(\mathbf{x}^l|\mathcal{H}_1^l\big)}{\sum_{\mathbf{x}^l} \exp\Big(-\frac{||\mathbf{y}^l - \sqrt{\rho^l}\mathbf{G}^l \mathbf{x}^l||^2}{\sigma^2_{e, l}}\Big) P\big(\mathbf{x}^l|\mathcal{H}_0^l\big)}\Bigg]
\end{align}
where, $\sigma^2_{e, l} \triangleq \sigma^2_{w, l} + \psi_l^2$, with $\sigma^2_{w, l}$ and $\psi_l^2$ as the power densities of the noise and interference processes respectively. In (\ref{eq5}), $\psi_l^2 = \psi^2_{l, \text{ICI}} + \psi^2_{l, \text{ISI}}$, where $\psi^2_{l, \text{ICI}}$ and $\psi^2_{l, \text{ISI}}$ are the ICI and ISI powers on the $l$th sub-carrier respectively.

It is worth mentioning here that practical implementation of the optimum rule in (\ref{eq5}) is severely difficult, as referred to in the case of WSNs in \cite{Ciuonzo2012} and \cite{ciuonzo2015}. It will be particularly problematic in case of collaborative WSS using multi-carrier massive MIMO DF, due to the lack of availability of $\mathbf{G}^l$, $P\big(\mathbf{x}^l|\mathcal{H}_i^l\big)$, $\sigma^2_{w, l}$ and $\psi_l^2$. The expression in (\ref{eq5}) is also numerically unstable due to the presence of exponential functions with large dynamics, especially for high SINR and/or large $K$. Hence we will resort to sub-optimum DF rules for WSS over multiple carrier frequency bands. They are easier to implement, require very little knowledge of the system parameters and offer numerical stability for realistic SINR values.

\subsection{Widely Linear (WL) Rules}\label{S4.2}

In this case, WL statistics is adopted, motivated by reduced complexity and $\mathbf{y}^l | \mathbf{G}^l, \mathcal{H}_i^l$ being an improper complex-valued random vector. Since the test statistics $\mathbf{\Gamma}_{i, l}^{\text{WL}}$ arises from WL processing of $\mathbf{y}^l$, we have
\begin{align} \label{eq13}
\mathbf{\Gamma}_{i, l}^{\text{WL}}|\mathbf{G}^l, \mathbf{x}^l \sim \mathcal{N}\big(\mathbb{E}\big\{\mathbf{\Gamma}_{i, l}^{\text{WL}}|\mathbf{G}^l, \mathbf{x}^l\big\}, \mathbb{V}\big\{\mathbf{\Gamma}_{i, l}^{\text{WL}}|\mathbf{G}^l, \mathbf{x}^l\big\}\big)
\end{align}
where $\mathbf{\Gamma}_{i,l}^{\text{WL}} \triangleq (\underline{\mathbf{a}}_{\text{WL},i}^l)^{\dagger} \underline{\mathbf{y}}^l$ and $\underline{\mathbf{a}}_{\text{WL},i}^l$ is chosen such that the deflection measure is maximized following, $\underline{\mathbf{a}}^l_{\text{WL}, i} \triangleq \text{max}_{\underline{\mathbf{a}}^l : ||\underline{\mathbf{a}}^l||^2}\mathcal{D}_i (\underline{\mathbf{a}}^l)$, where $\mathcal{D}_i (\underline{\mathbf{a}}^l) \triangleq (\mathbb{E}\{\mathbf{\Gamma}_{l}^{\text{WL}}|\mathcal{H}^l_1\} - \mathbb{E}\{\mathbf{\Gamma}_{l}^{\text{WL}}|\mathcal{H}^l_0\})^2/\mathbb{V}\{\mathbf{\Gamma}_{l}^{\text{WL}}|\mathcal{H}^l_i\}$, $\mathcal{D}_0 (\underline{\mathbf{a}}^l)$ and $\mathcal{D}_1 (\underline{\mathbf{a}}^l)$ correspond to the normal and modified deflections respectively \cite{quan2008}. The explicit expressions for $\underline{\mathbf{a}}^l_{\text{WL}, i}$ can be given by,
\begin{align} \label{eq14a}
\underline{\mathbf{a}}^l_{\text{WL}, i} = \frac{\mathbf{\Sigma}^{- 1}_{\underline{\mathbf{y}}^l|\mathbf{G}^l, \mathcal{H}^l_i} \underline{\mathbf{G}}^l \pmb{\mu}_i^l}{\Big|\Big|\mathbf{\Sigma}^{- 1}_{\underline{\mathbf{y}}^l|\mathbf{G}^l, \mathcal{H}^l_i} \underline{\mathbf{G}}^l \pmb{\mu}_i^l\Big|\Big|}
\end{align}
following the proposition made in \cite{ciuonzo2015}, where $\mathbf{\Sigma}_{\underline{\mathbf{y}}^l|\mathbf{G}^l, \mathcal{H}^l_i} = \big(\rho^l \underline{\mathbf{G}}^l \mathbf{\Sigma}_{\mathbf{x}^l|\mathcal{H}^l_i} (\underline{\mathbf{G}}^l)^{\dagger} + \sigma^2_{e, l} \mathbf{I}_{2N}\big)$ and $\pmb{\mu}_i^l \triangleq 2 \big[\big(P_{D, 1}^{l} - P_{F, 1}^{l}\big) \dotso \big(P_{D, K}^{l} - P_{F, K}^{l}\big)\big]^t$. The aforementioned expressions are based on the fact that the deflection-optimization is optimal only for a mean-shifted Gauss-Gauss hypothesis testing (i.e. $\mathbf{y}^l|\mathbf{G}^l, \mathcal{H}^l_i \sim \mathcal{N}_{\mathbb{C}}(\pmb{\mu}_i^l, \mathbf{\Sigma}_i^l)$), where normal and modified deflections coincide and they both represent the SINR of the statistics under Neyman-Pearson framework.
Additionally, although WL rules relax implementation requirements of the optimum rule, they still take into account the individual SU reliabilities via both terms $\mathbf{\Sigma}_{\mathbf{x}^{l}|\mathcal{H}_{i}^{l}}$ and $\pmb{\mu}_{i}^{l}$.

In order to derive the exact IS system probabilities for this fusion rule, we can deduce,
\begin{align} \label{eq14}
\mathbb{E}\big\{\mathbf{\Gamma}_{i, l}^{\text{WL}}|\mathbf{G}^l, \mathbf{x}^l\big\} &= \frac{\sqrt{\rho^l} (\pmb{\mu}_i^l)^t (\underline{\mathbf{G}}^l)^{\dagger} \mathbf{\Sigma}^{- 1}_{\underline{\mathbf{y}}^l|\mathbf{G}^l, \mathcal{H}^l_i} \underline{\mathbf{G}}^l \mathbf{x}^l}{||\mathbf{\Sigma}^{- 1}_{\underline{\mathbf{y}}^l|\mathbf{G}^l, \mathcal{H}^l_i} \underline{\mathbf{G}}^l \mathbf{x}^l||} \nonumber\\
\mathbb{V}\big\{\mathbf{\Gamma}_{i, l}^{\text{WL}}|\mathbf{G}^l, \mathbf{x}^l\big\} &= 2 \sigma^2_{e, l}.
\end{align}
It is apparent from (\ref{eq14}) that, $\mathbb{V}\big\{\mathbf{\Gamma}_{i, l}^{\text{WL}}|\mathbf{G}^l, \mathbf{x}^l\big\}$ does not depend on $\mathbf{x}^l$. Thus we can define $\mathbf{\Gamma}_{i, \text{WL}}^{l, \text{WB}} \triangleq \frac{\mathbf{\Gamma}_{i}^{l, \text{WL}}}{\sigma_{e, l}}$. Based on (\ref{eq14}), $\mathbf{\Gamma}_{i, \text{WL}}^{l, \text{WB}}|\mathbf{G}^l, \mathcal{H}^l_j$ is distributed as $\mathbf{\Gamma}_{i, \text{WL}}^{l, \text{WB}}|\mathbf{G}^l, \mathcal{H}^l_j \sim \sum_{\mathbf{x}^l \in {\mathbf{\chi}}^K} P\big(\mathbf{x}^l|\mathcal{H}_j^l\big) \mathcal{N} \big(\mathbb{E}\big\{\mathbf{\Gamma}_{i, \text{WL}}^{l, \text{WB}}|\mathbf{G}^l, \mathbf{x}^l\big\}, 1\big)$ where,
\begin{align} \label{eq15}
\mathbb{E}\big\{\mathbf{\Gamma}_{i, \text{WL}}^{l, \text{WB}}|\mathbf{G}^l, \mathbf{x}^l\big\} &= \frac{\sqrt{2\rho^l} (\pmb{\mu}_i^l)^t (\underline{\mathbf{G}}^l)^{\dagger} \mathbf{\Sigma}^{- 1}_{\underline{\mathbf{y}}^l|\mathbf{G}^l, \mathcal{H}^l_i} \underline{\mathbf{G}}^l \mathbf{x}^l}{\sigma_{e, l}||\mathbf{\Sigma}^{- 1}_{\underline{\mathbf{y}}^l|\mathbf{G}^l, \mathcal{H}^l_i} \underline{\mathbf{G}}^l \mathbf{x}^l||} 
\end{align}
For a large system, i.e. as $N \to \infty$, 
\begin{align} \label{eq16}
\lim_{N \to \infty}\big(\mathbb{E}\big\{\mathbf{\Gamma}_{i, \text{WL}}^{l, \text{WB}}|\mathbf{G}^l, \mathbf{x}^l\big\}\big) &= \frac{N \sqrt{2\rho^l} (\pmb{\mu}_i^l)^t \mathbf{x}^l \mathbf{V}_i^l \mathbf{D}_g^l}{\sigma_{e, l}\sqrt{(\pmb{\mu}_i^l)^t \mathbf{V}_i^l \mathbf{D}_g^l (\mathbf{V}_i^l)^t \pmb{\mu}_i^l}} 
\end{align}
where $\mathbf{V}_i^l \triangleq \mathbf{I}_K - \bigg(\frac{1 + \sigma_{w, l}^2 + \sum_{k = 1}^K {\overline{\mathbf{\tau}}}_{l, k}^2 + \sum_{k = 1}^K \sum_{p = 1}^{L} \big| \overline{\mathbf{\beta}}^{p}_k (d_{pl})\big|^2}{2 \mathbf{D}_g^l \rho^l N \sqrt{N} \mathbf{\Sigma}^{- 1}_{\mathbf{x}^l|\mathcal{H}^l_i}}\bigg)^{-1}$ for $l \neq p$ and $\mathbf{D}_g^l = \mathbf{D}*[\text{diag}(\mathcal{B}^l_k)]$. If $Z = K$, then $\mathbf{D}_g^l$ will be a diagonal matrix whose $k$th element equals $d^l_{g, k} = \lambda_k \beta^l_k (k -1)$. Equation (\ref{eq16}) denotes a mixture of real-valued Gaussians, all depending on $\mathbf{G}^l$ (which is random) through their mean. The combination of the noise and interference is also Gaussian as the interfering power is dependent on $\mathbf{H}^l$. 

If $Z = K$, then (\ref{eq16}) can be simplified in the case of conditionally uncorrelated decisions, $\mathbb{E}\{x^l_k, x^l_r | \mathcal{H}^l_j\} = \mathbb{E}\{x^l_k | \mathcal{H}^l_j\}\mathbb{E}\{x^l_r | \mathcal{H}^l_j\}~~(k \neq r)$ to obtain,
\begin{align} \label{eq18}
&\lim_{N \to \infty}\big(\mathbb{E}\big\{\mathbf{\Gamma}_{i, \text{WB}}^{l, \text{WL}}|\mathbf{G}^l, \mathbf{x}^l\big\}\big) \nonumber\\
&= \frac{N \sqrt{2\rho^l} \sum_{k =1}^K d^l_{g, k} \mu_{1, 0, k, l} x^l_k (\sigma_{e, l}^2 + 2\rho^l \sqrt{N}d^l_{g, k} \Sigma^{l, k}_{x})^{- 1}}{\sigma_{e, l}\sqrt{\sum_{k =1}^K d^l_{g, k} \mu^2_{1, 0, k, l} (\sigma_{e, l}^2 + 2\rho^l \sqrt{N}d^l_{g, k} \Sigma^{l, k}_{x})^{- 2}}} 
\end{align}
where $\Sigma^{l, k}_{x} = \mathbb{V}\{x_k^l | \mathcal{H}_i^l\}$, $d^l_{g, k} = \lambda_k \beta^l_k (k -1)$ and $\sigma_{e, l}^2 = \sigma_{w, l}^2 + \sum_{k = 1}^K {\overline{\mathbf{\tau}}}_{l, k}^2 + \sum_{k = 1}^K \sum_{p = 1}^{L} \big| \overline{\mathbf{\beta}}^{p}_k (d_{pl})\big|^2$.

Nonetheless, a large system approximation of the threshold level $\tilde{\gamma}^l$ for the $l$th sub-channel with reduced system knowledge can be found. Given a target $P_{F_0}^{l}$, the result can be stated using Proposition-\ref{S2p}, if the channel on each sub-carrier is characterized using number of channel taps equal to the number of SUs transmitting over each sub-carrier.
\begin{proposition}\label{S2p}
Assuming $\mathbb{E}\{\mathbf{x}^l | \mathcal{H}_0^l\} \triangleq (2 P_{F}^{l} - 1) \mathbf{1}_K$ and $\mathbb{E}\{(\mathbf{x}^l - \mathbb{E}\{\mathbf{x}^l | \mathcal{H}_0^l\})(\mathbf{x}^l - \mathbb{E}\{\mathbf{x}^l | \mathcal{H}_0^l\})^t| \mathcal{H}_0^l\} \triangleq [1 - (2 P_{F}^{l} - 1)^2] \mathbf{I}_K$, then a low-SINR large system $\tilde{\gamma}^l$ for approaching a target $\tilde{P}_{F_0}^{l}$ is given by,
\begin{align} \label{wl_rules}
\tilde{\gamma}^l \triangleq &~Q^{- 1}(\tilde{P}_{F_0}^{l}) \sqrt{2((1 - {\delta^l}^2) \cdot K + \sigma_{e, l}^2)} \nonumber\\
&+ \frac{2N \delta^l \sqrt{\rho^l} \sum_{k =1}^K \lambda_k \beta_k^l(k -1) \mu_{1, 0, k, l}}{\sqrt{\sum_{k =1}^K \lambda_k \beta_k^l(k -1) } \mu^2_{1, 0, k, l}}
\end{align}
where $\delta^l = (2 P_{F}^{l} - 1)$.
\end{proposition}
\subsubsection*{Proof}
See Appendix A.

\subsection{Maximal Ratio Combining (MRC) Rules}\label{S4.3}

The LLR in (\ref{eq5}) can be simplified under the assumption of perfect sensors, i.e. $P(\mathbf{x}^l = \mathbf{1}_K | \mathcal{H}_1^l) = P(\mathbf{x}^l = - \mathbf{1}_K | \mathcal{H}_0^l) = 1$. In this case, $\mathbf{x}^l \in \{\mathbf{1}_K, - \mathbf{1}_K\}$ and (\ref{eq5}) reduces to,
\begin{align} \label{eq20}
\ln \Bigg[\frac{\exp\Big(-\frac{||\mathbf{y}^l - \sqrt{\rho^l}\mathbf{G}^l \mathbf{1}_K||^2}{\sigma^2_{e, l}}\Big)}{\exp\Big(-\frac{||\mathbf{y}^l + \sqrt{\rho^l}\mathbf{G}^l \mathbf{1}_K||^2}{\sigma^2_{e, l}}\Big)}\Bigg] \propto \mathbb{R}\big\{\big(\mathbf{a}^l_{\text{MRC}}\big)^{\dagger}\mathbf{y}^l\big\} \triangleq \mathbf{\Gamma}^l_{\text{MRC}}
\end{align}
where, $\mathbf{a}^l_{\text{MRC}} \triangleq \mathbf{G}^l \mathbf{1}_K$ and terms independent of $\mathbf{y}^l$ have been incorporated in $\gamma^l$ as in (\ref{eq5}). In this case, we observe that $\mathbf{y}^l | \mathbf{G}^l, \mathbf{x}^l \sim \mathcal{N}_{\mathbb{C}} \big(\sqrt{\rho^l} \mathbf{G}^l \mathbf{x}^l, \sigma_{e, l}^2 \mathbf{I}_N\big)$. As an immediate consequence, we have $\mathbf{\Gamma}^l_{\text{MRC}} | \mathbf{G}^l, \mathbf{x}^l \sim \mathcal{N} \big(\mathbb{E}\{\mathbf{\Gamma}^l_{\text{MRC}} | \mathbf{G}^l, \mathbf{x}^l\}, \mathbb{V}\{\mathbf{\Gamma}^l_{\text{MRC}} | \mathbf{G}^l, \mathbf{x}^l\}\big)$ after MRC processing of $\mathbf{y}^l$.
In this case also it can be shown that, $\mathbb{V}\{\mathbf{\Gamma}^l_{\text{MRC}} | \mathbf{G}^l, \mathbf{x}^l\}$ does not depend on $\mathbf{x}^l$. Thus we can define, $\mathbf{\Gamma}_{\text{MRC}}^{l, \text{WB}} \triangleq \frac{\sqrt{2} \mathbf{\Gamma}^l_{\text{MRC}}}{\sigma_{e, l}~||\mathbf{a}^l_{\text{MRC}}||}$ and evaluate the performance in terms of $\mathbf{\Gamma}_{\text{MRC}}^{l, \text{WB}}$. In that case, $\big\{\mathbf{\Gamma}_{\text{MRC}}^{l, \text{WB}} | \mathbf{G}^l \mathcal{H}_j^l\big\}$ is distributed as, $\mathbf{\Gamma}_{\text{MRC}}^{l, \text{WB}} | \mathbf{G}^l \mathcal{H}_j^l \sim \sum_{\mathbf{x}^l \in {\mathbf{\chi}}^K} P\big(\mathbf{x}^l|\mathcal{H}_j^l\big) \mathcal{N} \big(\mathbb{E}\big\{\mathbf{\Gamma}_{\text{MRC}}^{l, \text{WB}}|\mathbf{G}^l, \mathbf{x}^l\big\}, 1\big)$, where, for a large system, i.e. as $N \to \infty$, 
\begin{align} \label{eq23}
\lim_{N \to \infty}\big(\mathbb{E}\big\{\mathbf{\Gamma}_{\text{MRC}}^{l, \text{WB}}|\mathbf{G}^l, \mathbf{x}^l\big\}\big) &= \frac{\sqrt{2 N \rho^l}~\mathbb{R}\{(\mathbf{1}_K)^t \mathbf{D}_g^l \mathbf{x}^l\}}{\sigma_{e, l}\sqrt{(\mathbf{1}_K)^t \mathbf{D}_g^l \mathbf{1}_K}} 
\end{align}

Additionally, in order to exploit the linear SINR increases with $N$, which would inevitably make the fusion process mainly dependent on the `sensing' errors (and consequently MRC rule becomes clearly in appropriate, since its design is unaware of sensing errors), we resort to an alternative form of MRC, denoted as modified MRC (mMRC) given by, $\mathbf{\Gamma}_{\text{mMRC}}^{l} \triangleq \mathbb{R}\big\{\big(\mathbf{a}^l_{\text{mMRC}}\big)^{\dagger}\mathbf{y}^l\big\}$ where, $\mathbf{a}^l_{\text{mMRC}} \triangleq \mathbf{G}^l (\mathbf{D}_g^l)^{- 1}\mathbf{1}_K$. In this case also, $\mathbb{V}\{\mathbf{\Gamma}_{\text{mMRC}}^{l} | \mathbf{G}^l, \mathbf{x}^l\}$ does not depend on $\mathbf{x}^l$. Thus, we define, $\mathbf{\Gamma}_{\text{mMRC}}^{l, \text{WB}} \triangleq \frac{\sqrt{2} \mathbf{\Gamma}_{\text{mMRC}}^{l}}{\sigma_{e, l}~||\mathbf{a}^l_{\text{mMRC}}||}$ and evaluate the performance in terms of $\mathbf{\Gamma}_{\text{mMRC}}^{l, \text{WB}}$ as,
\begin{align} \label{eq25}
\lim_{N \to \infty}\big(\mathbb{E}\big\{\mathbf{\Gamma}_{\text{mMRC}}^{l, \text{WB}}|\mathbf{G}^l, \mathbf{x}^l\big\}\big) &= \frac{\sqrt{2 N \rho^l}~\mathbb{R}\{(\mathbf{1}_K)^t \mathbf{x}^l\}}{\sigma_{e, l}\sqrt{(\mathbf{1}_K)^t (\mathbf{D}_g^l)^{- 1} \mathbf{1}_K}} 
\end{align}
for a large system. Following the same argument as in the case of (\ref{eq16}), (\ref{eq23}) and (\ref{eq25}) denote mixtures of real-valued Gaussians. 

If $Z = K$, then the MRC and mMRC rules can be simplified in the case of conditionally uncorrelated decisions to obtain,
\begin{align} \label{eq27}
\lim_{N \to \infty}\big(\mathbb{E}\big\{\mathbf{\Gamma}_{\text{MRC}}^{l, \text{WB}}|\mathbf{G}^l, \mathbf{x}^l\big\}\big) &= \frac{\sqrt{2 N \rho^l} \sum_{k =1}^K d^l_{g, k} x^l_k}{\sigma_{e, l}\sqrt{\sum_{k =1}^K d^l_{g, k}}} \\
\lim_{N \to \infty}\big(\mathbb{E}\big\{\mathbf{\Gamma}_{\text{mMRC}}^{l, \text{WB}}|\mathbf{G}^l, \mathbf{x}^l\big\}\big) &= \frac{\sqrt{2 N \rho^l} \sum_{k =1}^K x^l_k}{\sigma_{e, l}\sqrt{\sum_{k =1}^K (d^l_{g, k})^{- 1}}} 
\end{align}
where $d^l_{g, k} = \lambda_k \beta^l_k (k -1)$ and $\sigma_{e, l}^2 = \sigma_{w, l}^2 + \sum_{k = 1}^K {\overline{\mathbf{\tau}}}_{l, k}^2 + \sum_{k = 1}^K \sum_{p = 1}^{L} \big| \overline{\mathbf{\beta}}^{p}_k (d_{pl})\big|^2$. 

\subsection{Time-Reversal (TR) based Fusion Rules}\label{S4.4}

When conventional DF rules are used, a residual interference remains and the SINR saturates at a certain level even for an infinite number of DFC antennas. This is due to the correlation between the combiner taps and the ISI and ICI components. In this section, we propose application of time-reversal (TR) methods to alleviate this saturation problem. Recently, academia has concentrated their effort to the application of TR for the future generation of wireless networks \cite{chen2016time}, especially to massive MIMO in the context of single-carrier transmission \cite{pitarokoilis2012optimality}. It is established in \cite{pitarokoilis2012optimality} that channel distortions tend to fade away as the number of BS antennas goes to infinity. In order to exploit advantages of TR methods when applied to large array regime, we propose application of TR-WL and TR-MRC fusion rules in collaborative WSS. 

\subsubsection{TR-WL Rule}\label{S4.4.2}

The first approach consists of adopting the TR variant of the WL statistics such that, $\mathbf{\Gamma}_{i,l}^{\text{TR-WL}} \triangleq (\underline{\mathbf{a}}_{\text{TR-WL},i}^l)^{\dagger} \underline{\mathbf{y}}^l$ where $\underline{\mathbf{a}}_{\text{TR-WL},i}^l$ can be explicitly expressed as,
\begin{align} \label{eq140}
\underline{\mathbf{a}}^l_{\text{TR-WL}, i} = \frac{\mathbf{\Sigma}^{- 1}_{\underline{\mathbf{y}}^l|\breve{\mathbf{G}}^l, \mathcal{H}^l_i} \underline{\breve{\mathbf{G}}}^l \pmb{\mu}_i^l}{\Big|\Big|\mathbf{\Sigma}^{- 1}_{\underline{\mathbf{y}}^l|\breve{\mathbf{G}}^l, \mathcal{H}^l_i} \underline{\breve{\mathbf{G}}}^l \pmb{\mu}_i^l\Big|\Big|}
\end{align}
following the formulation and proposition made in Subsection~\ref{S4.1}, where $\mathbf{\Sigma}_{\underline{\mathbf{y}}^l|\breve{\mathbf{G}}^l, \mathcal{H}^l_i} = \big(\rho^l \underline{\breve{\mathbf{G}}}^l \mathbf{\Sigma}_{\mathbf{x}^l|\mathcal{H}^l_i} (\underline{\breve{\mathbf{G}}}^l)^{\dagger} + \sigma^2_{e, l} \mathbf{I}_{2N}\big)$. Using the definition, $\mathbf{\Gamma}_{i, \text{TR-WL}}^{l, \text{WB}} \triangleq \frac{\mathbf{\Gamma}_{i}^{l, \text{TR-WL}}}{\sigma_{e, l}}$ and the test statistics $\mathbf{\Gamma}_{i}^{l,\text{TR-WL}}$ being distributed as $\mathbf{\Gamma}_{i}^{l, \text{TR-WL}}|\mathbf{G}^l,\mathbf{x}^l \sim \mathcal{N}(\mathbb{E}\{\mathbf{\Gamma}_{i}^{l, \text{TR-WL}}|\mathbf{G}^l,\mathbf{x}^l\}, \mathbb{V}\{\mathbf{\Gamma}_{i}^{l, \text{TR-WL}}|\mathbf{G}^l,\mathbf{x}^l\}),~\mathbf{\Gamma}_{i, \text{TR-WL}}^{l, \text{WB}}|\mathbf{G}^l, \mathcal{H}^l_j$ will be distributed as $\mathbf{\Gamma}_{i, \text{TR-WL}}^{l, \text{WB}}|\mathbf{G}^l, \mathcal{H}^l_j \sim \sum_{\mathbf{x}^l \in {\mathbf{\chi}}^K} P\big(\mathbf{x}^l|\mathcal{H}_j^l\big) \mathcal{N} \big(\mathbb{E}\big\{\mathbf{\Gamma}_{i, \text{TR-WL}}^{l, \text{WB}}|\mathbf{G}^l, \mathbf{x}^l\big\}, 1\big)$. Here,
\begin{align} \label{eq141}
\mathbb{E}\big\{\mathbf{\Gamma}_{i, \text{TR-WL}}^{l, \text{WB}}|\mathbf{G}^l, \mathbf{x}^l\big\} &= \frac{\sqrt{2\rho^l} (\pmb{\mu}_i^l)^t (\underline{\mathbf{G}}^l)^{\dagger} \mathbf{\Sigma}^{- 1}_{\underline{\mathbf{y}}^l|\breve{\mathbf{G}}^l, \mathcal{H}^l_i} \underline{\breve{\mathbf{G}}}^l \mathbf{x}^l}{\sigma_{e, l}||\mathbf{\Sigma}^{- 1}_{\underline{\mathbf{y}}^l|\breve{\mathbf{G}}^l, \mathcal{H}^l_i} \underline{\breve{\mathbf{G}}}^l \mathbf{x}^l||} 
\end{align}
As $N \to \infty$, we have,
\begin{align} \label{eq142}
\lim_{N \to \infty}\big(\mathbb{E}\big\{\mathbf{\Gamma}_{i, \text{WL}}^{l, \text{WB}}|\mathbf{G}^l, \mathbf{x}^l\big\}\big) &= \frac{N \sqrt{2\rho^l} (\pmb{\mu}_i^l)^t \mathbf{x}^l \breve{\mathbf{V}}_i^l \mathbf{F}^l}{\sigma_{e, l}\sqrt{(\pmb{\mu}_i^l)^t \breve{\mathbf{V}}_i^l \breve{\mathbf{A}}^l (\breve{\mathbf{V}}_i^l)^t \pmb{\mu}_i^l}} 
\end{align}
where $\breve{\mathbf{V}}_i^l \triangleq \mathbf{I}_K - \bigg(\frac{1 + \sigma_{w, l}^2 + \sum_{k = 1}^K {\overline{\mathbf{\tau}}}_{l, k}^2 + \sum_{k = 1}^K \sum_{p = 1}^{L} \big| \overline{\mathbf{\beta}}^{p}_k (d_{pl})\big|^2}{2 \breve{\mathbf{A}}^l \rho^l N \sqrt{N} \mathbf{\Sigma}^{- 1}_{\mathbf{x}^l|\mathcal{H}^l_i}}\bigg)^{-1}$ for $l \neq p$. If $Z = K$, then $\mathbf{F}^l$ will be a diagonal matrix whose $k$th element equals to $\lambda_k \beta^l_k (K - k) \beta^l_k (k - 1)$ and $\breve{\mathbf{A}}$ will be another diagonal matrix with $k$th element $\lambda_k \beta^l_k (K - k)$. Hence, for the case of $Z = K$, (\ref{eq142}) can be simplified in the case of conditionally uncorrelated decisions to obtain,
\begin{align} \label{eq143}
&\lim_{N \to \infty}\big(\mathbb{E}\big\{\mathbf{\Gamma}_{i, \text{WB}}^{l, \text{TR-WL}}|\mathbf{G}^l, \mathbf{x}^l\big\}\big) \nonumber\\
&= \frac{N \sqrt{2\rho^l} \sum_{k =1}^K \lambda_k^l \sqrt{\beta_k^l (K - k)} \sqrt{\beta_k^l (k - 1)} \mu_{1, 0, k, l} x^l_k (\breve{v}_{i,k}^l)^{- 1}}{\sigma_{e, l}\sqrt{\sum_{k =1}^K \lambda_k^2 {\beta^l_k}^2(K - k) \mu^2_{1, 0, k, l} (\breve{v}_{i,k}^l)^{- 2}}} 
\end{align}
where $\breve{v}_{i,k}^l = \sigma_{e, l}^2 + 2\rho^l \sqrt{N}\lambda_k \beta^l_k (K - k) \Sigma^{l, k}_{x}$ and $\Sigma^{l, k}_{x} = \mathbb{V}\{x_k^l | \mathcal{H}_i^l\}$.

Assuming $\mathbb{E}\{\mathbf{x}^l | \mathcal{H}_0^l\} \triangleq (2 P_{F}^{l} - 1) \mathbf{1}_K$ and $\mathbb{E}\{(\mathbf{x}^l - \mathbb{E}\{\mathbf{x}^l | \mathcal{H}_0^l\})(\mathbf{x}^l - \mathbb{E}\{\mathbf{x}^l | \mathcal{H}_0^l\})^t| \mathcal{H}_0^l\} \triangleq [1 - (2 P_{F}^{l} - 1)^2] \mathbf{I}_K$, then a low-SINR large system $\tilde{\gamma}^l_{\text{TR-WL}}$ for approaching a target $\tilde{P}_{F_0}^{l}$ is given by,
\begin{align} \label{eq19}
&\tilde{\gamma}^l_{\text{TR-WL}} \triangleq ~Q^{- 1}(\tilde{P}_{F_0}^{l}) \sqrt{2((1 - {\delta^l}^2) \cdot K + \sigma_{e, l}^2)} \nonumber\\
&\quad + \frac{2N \delta^l \sqrt{\rho^l} \sum_{k =1}^K \lambda_k^l \sqrt{\beta_k^l (K - k)} \sqrt{\beta_k^l (k - 1)} \mu_{1, 0, k, l}}{\sqrt{\sum_{k =1}^K \lambda_k \beta_k^l(K - k) } \mu^2_{1, 0, k, l}}
\end{align}
where $\delta^l = (2 P_{F}^{l} - 1)$.

\subsubsection{(Modified) TR-MRC Rules}\label{S4.4.3}

The TR-MRC rule can be defined as, $\mathbf{a}^l_{\text{TR-MRC}} \triangleq \breve{\mathbf{G}}^l \mathbf{1}_K$, and the test statistics as, $\mathbf{\Gamma}^{l, \text{WB}}_{\text{TR-MRC}} \triangleq \frac{\sqrt{2} \mathbf{\Gamma}^{l}_{\text{TR-MRC}}}{\sigma_{e, l}~||\mathbf{a}^l_{\text{TR-MRC}}||}$ and,
\begin{align} \label{eq31}
\mathbb{E}\big\{\mathbf{\Gamma}^{l, \text{WB}}_{i, \text{TR-MRC}}|\mathbf{G}^l, \mathbf{x}^l\big\} &= \frac{\sqrt{2 N \rho^l}~\mathbb{R}\{(\mathbf{1}_K)^t \mathbf{F}^l \mathbf{x}^l\}}{\sigma_{e, l}\sqrt{(\mathbf{1}_K)^t \breve{\mathbf{A}}^l \mathbf{1}_K}} 
\end{align}
for $N \to \infty$. In order to exploit the linear SINR increase with $N$, we device an alternative form of mMRC, denoted as time-reversed modified MRC (TR-mMRC) given by, $\mathbf{\Gamma}^{l}_{\text{TR-mMRC}} \triangleq \mathbb{R}\big\{\big(\mathbf{a}^l_{\text{TR-mMRC}}\big)^{\dagger}\mathbf{y}^l\big\}$ where, $\mathbf{a}^l_{\text{TR-mMRC}} \triangleq \breve{\mathbf{G}}^l ({\breve{\mathbf{A}}}^l)^{- 1}\mathbf{1}_K$. Thus we can define, $\mathbf{\Gamma}^{l, \text{WB}}_{\text{TR-mMRC}} \triangleq \frac{\sqrt{2} \mathbf{\Gamma}^{l}_{\text{TR-mMRC}}}{\sigma_{e, l}~||\mathbf{a}^l_{\text{TR-mMRC}}||}$ and, in case of $N \in \infty$, evaluate the performance in terms of $\mathbf{\Gamma}^{l, \text{WB}}_{\text{TR-mMRC}}$ as,
\begin{align} \label{eq33}
&\mathbb{E}\big\{\mathbf{\Gamma}^{l, \text{WB}}_{i, \text{TR-mMRC}}|\mathbf{G}^l, \mathbf{x}^l\big\} \nonumber\\
&= \frac{\sqrt{2 N \rho^l}~\mathbb{R}\{(\mathbf{1}_K)^t (\breve{\mathbf{G}}^l)^{\dagger} (({\breve{\mathbf{A}}}^l)^{- 1})^{\dagger} \mathbf{G}^l \mathbf{x}^l\}}{\sigma_{e, l}\sqrt{(\mathbf{1}_K)^t  (({\breve{\mathbf{A}}}^l)^{- 1})^{\dagger} \mathbf{1}_K}} 
\end{align}

If $Z = K$, then the TR-MRC and TR-mMRC rules can be simplified in the case of conditionally uncorrelated decisions and large $N$ at the DFC to obtain,
\begin{align} \label{eq34}
&\mathbb{E}\big\{\mathbf{\Gamma}^{l, \text{WB}}_{i, \text{TR-MRC}}|\mathbf{G}^l, \mathbf{x}^l\big\} \nonumber\\
&\quad= \frac{ \sqrt{2 N \rho^l} \sum_{k =1}^K x^l_k \lambda_k^l \sqrt{\beta_k^l (K - k)} \sqrt{\beta_k^l (k - 1)}}{\sigma_{e, l}\sqrt{\sum_{k =1}^K \lambda_k^l \beta_k^l (K - k)}} \\
&\mathbb{E}\big\{\mathbf{\Gamma}^{l, \text{WB}}_{i, \text{TR-mMRC}}|\mathbf{G}^l, \mathbf{x}^l\big\} \nonumber\\
&= \frac{\sqrt{2 N \rho^l} \sum_{k =1}^K x^l_k \sqrt{\beta_k^l (k - 1)}/\sqrt{\beta_k^l (K - k)}}{\sigma_{e, l}\sqrt{\sum_{k =1}^K (\lambda_k^l \beta^l_k (K - k))^{- 1}}} 
\end{align}
where $\sigma_{e, l}^2 = \sigma_{w, l}^2 + \sum_{k = 1}^K {\overline{\mathbf{\tau}}}_{l, k}^2 + \sum_{k = 1}^K \sum_{p = 1}^{L} \big| \overline{\mathbf{\beta}}^{p}_k (d_{pl})\big|^2$.

In the following proposition, we derive large system approximation of the threshold level $\tilde{\gamma}^l$ for the $l$th sub-channel with reduced system knowledge. For decision fusion, we consider the set of MRC and TR-MRC rules presented in Subsections~\ref{S4.2} and \ref{S4.3}. 

\begin{proposition}\label{S5p}
Given a target $P_{F_0}^{l}$ in the case of (conditionally) uncorrelated sensor decisions with $P_{F, k}^{l} = P_{F}^{l}, k \in \mathcal{K}$, and assuming $\mathbb{E}\{\mathbf{x}^l | \mathcal{H}_0^l\} \triangleq (2 P_{F}^{l} - 1) \mathbf{1}_K$ and $\mathbb{E}\{(\mathbf{x}^l - \mathbb{E}\{\mathbf{x}^l | \mathcal{H}_0^l\})(\mathbf{x}^l - \mathbb{E}\{\mathbf{x}^l | \mathcal{H}_0^l\})^t| \mathcal{H}_0^l\} \triangleq [1 - (2 P_{F}^{l} - 1)^2] \mathbf{I}_K$, then a low-SINR large system $\tilde{\gamma}^l$ for approaching a target ${P}_{F_0}^{l}$ using the above-mentioned set of MRC rules is given by,
\begin{align} \label{eq35}
\tilde{\gamma}^l_{\text{MRC}} &\triangleq~Q^{- 1}({P}_{F_0}^{l}) \sqrt {\frac{\alpha^l}{2}}\sum_{k =1}^K d_{g, k}^l + \delta^l \sqrt{N \rho^l}\sum_{k =1}^K \big(d_{g, k}^l\big)^2 \nonumber\\
\tilde{\gamma}^l_{\text{TR-MRC}} &\triangleq~Q^{- 1}({P}_{F_0}^{l}) \sqrt {\frac{\alpha^l}{2}}\sum_{k =1}^K \breve{g}_k^l + \delta^l \sqrt{N \rho^l}\sum_{k =1}^K \big(\breve{g}_k^l\big)^2 \nonumber\\
\tilde{\gamma}^l_{\text{TR-mMRC}} &\triangleq~Q^{- 1}({P}_{F_0}^{l}) \sqrt {\frac{\alpha^l}{2}}\sum_{k =1}^K d_{g, k}^l + \delta^l \sqrt{N \rho^l}\sum_{k =1}^K \big(\breve{g}_k^l\big)^{-2}
\end{align}
where $g_k^l = \lambda_k^l \beta_k^l(k -1)$, $\breve{g}_k^l = \lambda_k^l \beta_k^l(K - k)$, $\alpha^l = ((1 - {\delta^l}^2)K + \sigma_{e, l}^2)$ and $\delta^l = (2 P_{F}^{l} - 1)$.
\end{proposition}
\subsubsection*{Proof}
See Appendix A.

\section{Performance Analysis} \label{S5}

\subsection{Performance Measures}\label{S5.1}

Combining the decisions from all the $K$ SUs independently on each sub-carrier, we can arrive at the total probabilities $P_{D_0}^{l}$ and $P_{F_0}^{l}$ for the network for our chosen fusion algorithms. Here, we compare the performance for different decision fusion rules both in terms of instantaneous sub-carrier (IS) system false alarm and detection probabilities defined as,
\begin{align} \label{eq11}
P_{F_0}^{l} (\gamma^l, \mathbf{G}^l) &\triangleq \text{Pr} \big\{\mathbf{\Gamma}^l > \gamma^l | \mathbf{G}^l, \mathcal{H}_0^l \big\} \nonumber\\
P_{D_0}^{l} (\gamma^l, \mathbf{G}^l) &\triangleq \text{Pr} \big\{\mathbf{\Gamma}^l > \gamma^l | \mathbf{G}^l, \mathcal{H}_1^l \big\}
\end{align}
and the corresponding sub-carrier average (SA) counter-parts,
\begin{align} \label{eq12}
P_{F_0}^{l} (\gamma^l) &\triangleq \mathbb{E}_{\mathbf{G}^l} \big\{P_{F_0}^{l} (\gamma^l, \mathbf{G}^l)\big\} = \text{Pr} \big\{\mathbf{\Gamma}^l > \gamma^l | \mathbf{G}^l, \mathcal{H}_0^l \big\} \nonumber\\
P_{D_0}^{l} (\gamma^l) &\triangleq \mathbb{E}_{\mathbf{G}^l} \big\{P_{D_0}^{l} (\gamma^l, \mathbf{G}^l)\big\} = \text{Pr} \big\{\mathbf{\Gamma}^l > \gamma^l | \mathbf{G}^l, \mathcal{H}_1^l \big\}
\end{align}
where $\mathbf{\Gamma}^l$ is the generic statistic employed at the DFC over the $l$th sub-carrier.

We highlight that the probability of detection $P_{D_0}^{l}$ should be high as it indicates the level of protection of the PU from the interfering SUs. On the other hand, low values of the false-alarm probability are necessary in order to maintain high opportunistic throughput, since false alarm events will prevent the unused frequency bands from being used by the SUs. Therefore, the choice of the threshold $\gamma^l$ leads to a trade-off between $P_{F_0}^{l}$ and the probability of missing a chance. Specifically, a higher threshold will result in a smaller $P_{F_0}^{l}$ and a larger probability of miss and vice-versa.
\subsection{Simulation Setup} \label{S5.set}

For simulating the performance of the set of fusion rules proposed herein, we assume (for simplicity) that the PU sensing process from SUs is based on conditionally independently and identically distributed (iid) decisions over the SUs and the sub-carriers, with $(P_{D, k}^l,P_{D, k}^l)=(P_D, P_F) = (0.5, 0.01)$.
Additionally, the SUs are located in a circular area around the DFC with radius $r_{\text{max}} = 1000$ m uniformly at random and we assume that no SU is closer to the DFC than $r_{\text{min}} = 100$ m.
In other terms, $r_{\text{min}} \leq r_k \leq r_{\text{max}}$, where $r_k$ is the distance between the $k$th SU and the DFC.
The large-scale shadowing between $k$th SU and the DFC (at $l$th sub-carrier) is modeled using $\lambda_k^l = \psi_k (\frac{r_{\text{min}}}{r_k})^n$, where $n$ denotes the path-loss exponent and $\psi_k$ is a log-normal random variable, i.e. $10 \log_{10}(\psi_k) \sim \mathcal{N}(\mu_{\lambda}, \sigma_{\lambda}^2)$, being $\mu_{\lambda}$ and $\sigma_{\lambda}$ the mean and standard deviation in dB respectively.
Finally, for simplicity, we set $\rho^l = 1/\sqrt{N}$ and $\sigma^2_{w, l} = 1$, and model the block-fading channel on each sub-carrier between the SUs and the DFC with channel taps equal in number to the number of SUs present in the network ($Z = K$).

\subsection{Numerical Results} \label{S5.2}

\begin{figure}[t]
\vspace*{-5mm}
\begin{center}
 \includegraphics[width=1.9\linewidth]{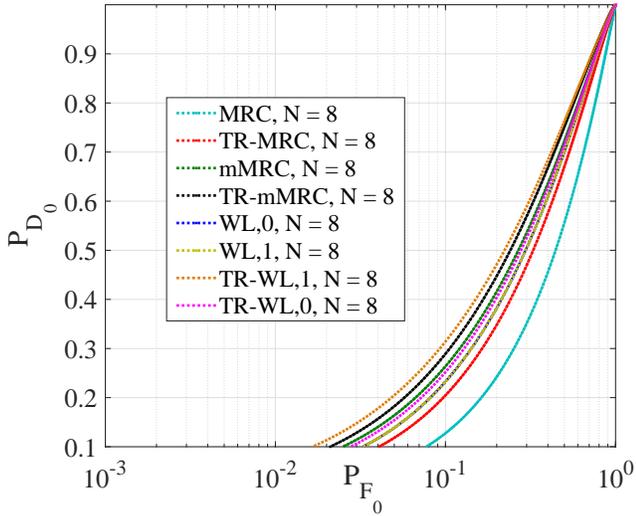}
\end{center}
\vspace*{-6mm}
\caption{$P_{D_0}$ vs. $P_{F_0}$ for all the presented set of MRC (MRC, mMRC, TR-MRC, TR-mMRC) and WL (WL, TR-WL) rules with $(\mu_{\lambda}, \sigma_{\lambda}) = (4, 2)$ dB and a path-loss exponent of 2, with moderate $N = 8$ and 8 SUs in the network.}
\label{FIG1a}
\end{figure}

\begin{figure}[t]
\vspace*{-5mm}
\begin{center}
 \includegraphics[width=1.9\linewidth]{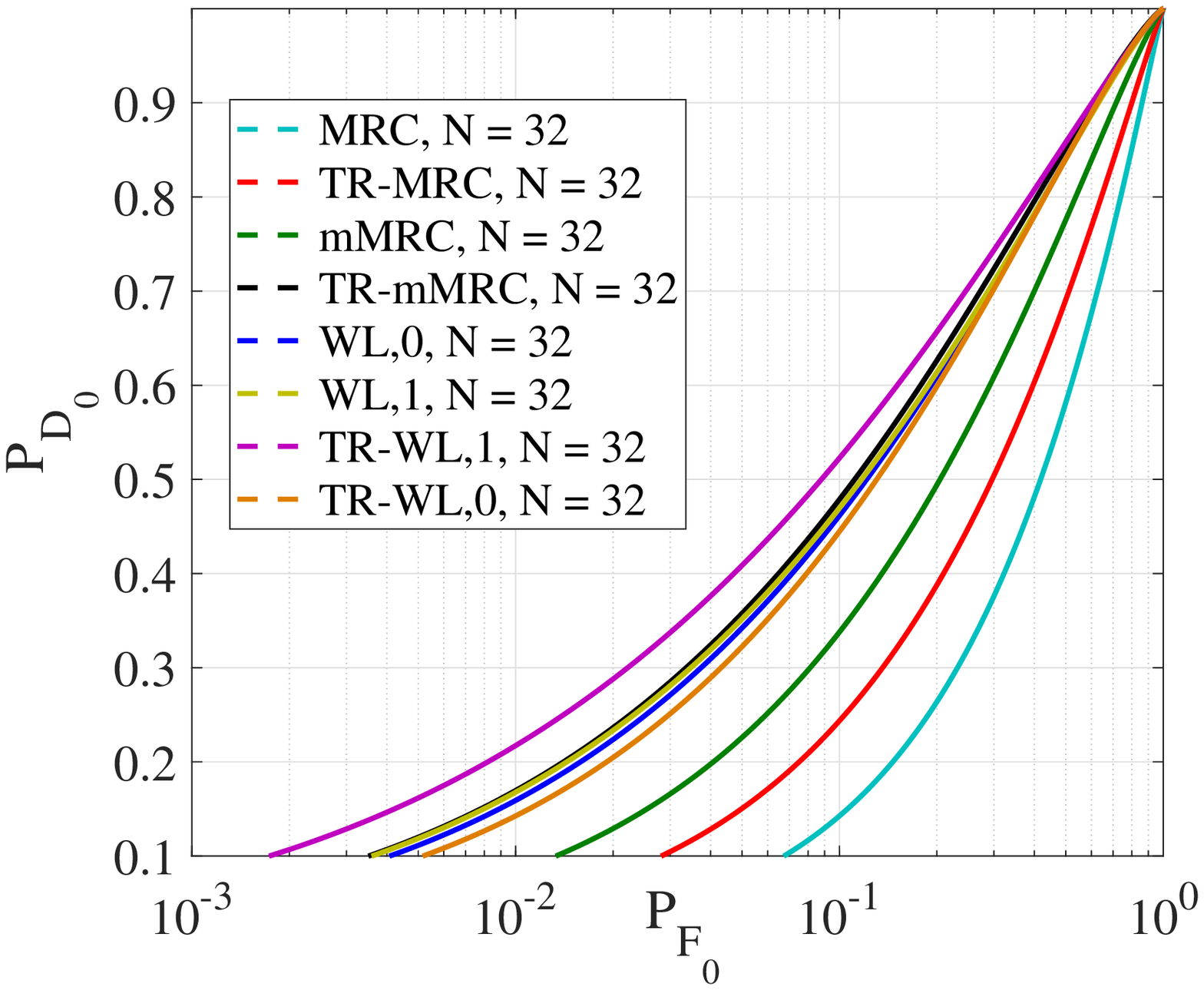}
\end{center}
\vspace*{-6mm}
\caption{$P_{D_0}$ vs. $P_{F_0}$ for all the presented set of MRC (MRC, mMRC, TR-MRC, TR-mMRC) and WL (WL, TR-WL) rules with $(\mu_{\lambda}, \sigma_{\lambda}) = (4, 2)$ dB and a path-loss exponent of 2, with large $N = 32$ and 8 SUs in the network.}
\label{FIG1b}
\end{figure}

In Fig.~\ref{FIG1a} and Fig.~\ref{FIG1b}, we show $P_{D_0}$ as a function of $P_{F_0}$ for two scenarios generated by varying $N$ with fixed $K = 8$. In these two figures, we consider a moderate $N = 8$ and large sized array $N = 32$ at the DFC. Contrary to the observations in \cite{ciuonzo2015}, in case of moderately large $N$, mMRC does not outperform MRC. For large $N$, both TR-MRC and TR-mMRC outperform MRC and mMRC. It is also apparent that when $N$ is moderate, TR variants of WL and MRC become more appealing solutions than ordinary MRC or WL. However, TR-WL offers better performance than TR-mMRC (refer to Fig.~\ref{FIG1a} and Fig.~\ref{FIG1b}). 

In Fig.~\ref{FIG1a}, WL rules perform quite close to TR based rules for moderately high $N = 32$ at the DFC, but suffers loss in performance with small $N$ in Fig.~\ref{FIG1b}. This can be owing to the reduced system knowledge available and introduction of interference due to channel impairments. It is to be mentioned here again that we are comparing performance from the context that frequency bands in the spectrum are closely spaced and no CP is used as a part of data transmission. It is also demonstrated that WL,1 performs slightly better than WL,0, as is shown in \cite{ciuonzo2015}.

\begin{figure}[t]
\begin{center}
 \includegraphics[width=1.9\linewidth]{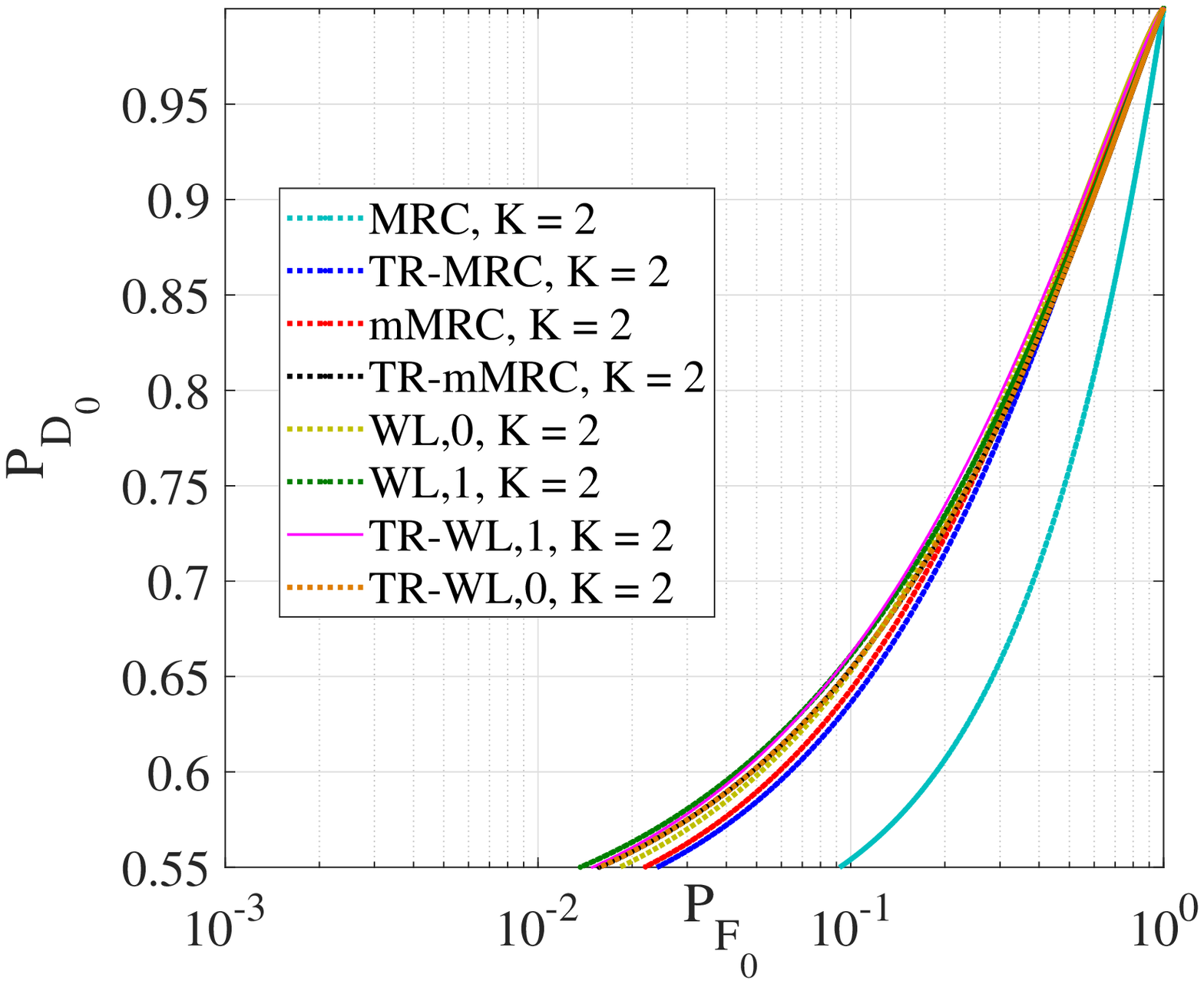}
\end{center}
\vspace*{-6mm}
\caption{$P_{D_0}$ vs. $P_{F_0}$ for all the presented set of MRC (MRC, mMRC, TR-MRC, TR-mMRC) and WL (WL, TR-WL) rules with $(\mu_{\lambda}, \sigma_{\lambda}) = (4, 2)$ dB and a path-loss exponent of 2, with small $K = 2$ and $N = 64$ at the DFC.}
\label{FIG2a}
\end{figure} 

\begin{figure}[t]
\begin{center}
 \includegraphics[width=1.9\linewidth]{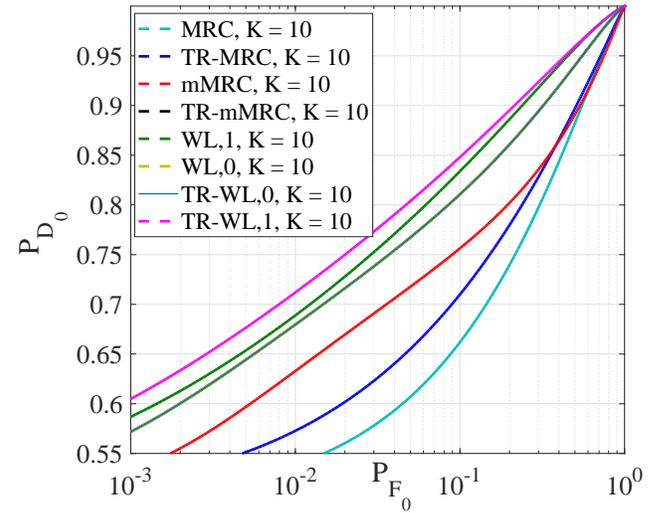}
\end{center}
\vspace*{-6mm}
\caption{$P_{D_0}$ vs. $P_{F_0}$ for all the presented set of MRC (MRC, mMRC, TR-MRC, TR-mMRC) and WL (WL, TR-WL) rules with $(\mu_{\lambda}, \sigma_{\lambda}) = (4, 2)$ dB and a path-loss exponent of 2, with moderate $K = 10$ and $N = 64$ at the DFC.}
\label{FIG2b}
\end{figure} 

In Fig.~\ref{FIG2a} and Fig.~\ref{FIG2b}, we consider a small $K = 2$ and moderate sized number of SUs $K = 2$ and moderate sized number of SUs $K = 10$ competing for the spectrum. In this case, the DFC is equipped with $N = 64$ antennas. It is evident that, for a small number of SUs, time-reversed versions offer no improvements over the ordinary techniques except MRC only. It is also evident that all the fusion rules exploit effectively the dramatic increase in diversity. However, enhancement in performance is observed by a considerable amount in case of TR based rules as soon as we have a large number of SUs competing for the spectrum. A major contributing factor for this improvement can be the fact that $K$ SUs collaborate to increase spatial diversity and time reversal of the channel matrix combats residual interference.

\begin{figure}[t]
\vspace*{-5mm}
\begin{center}
 \includegraphics[width=1.9\linewidth]{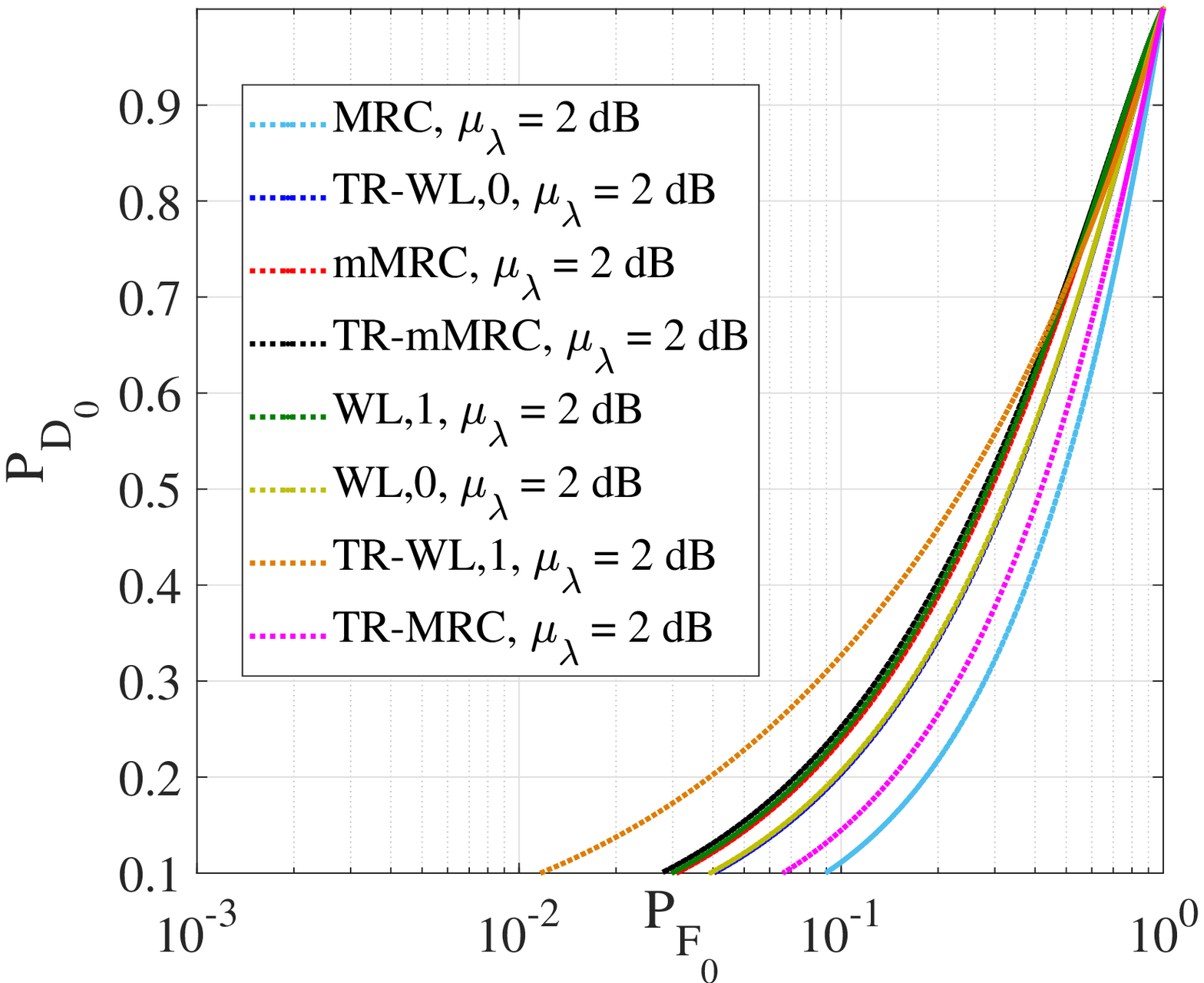}
\end{center}
\vspace*{-6mm}
\caption{$P_{D_0}$ vs. $P_{F_0}$ for all the presented set of MRC (MRC, mMRC, TR-MRC, TR-mMRC) and WL (WL, TR-WL) rules with $(N, K) = (32, 8)$ with mean of shadowing effect $\mu_{\lambda} = 2$ dB (indoor) and a path-loss exponent of 2 in the network.}
\label{FIG3a}
\end{figure} 

\begin{figure}[t]
\vspace*{-5mm}
\begin{center}
 \includegraphics[width=1.9\linewidth]{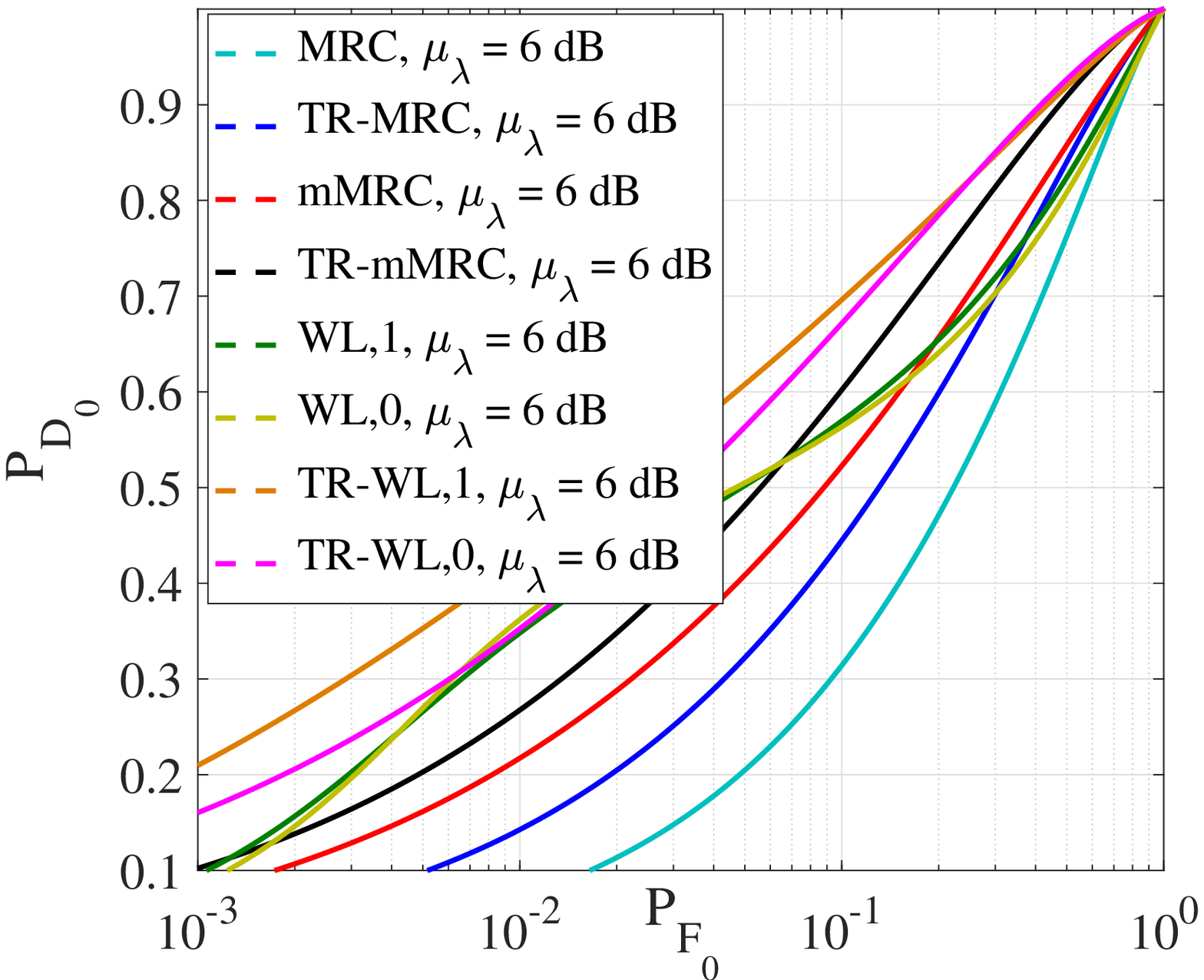}
\end{center}
\vspace*{-6mm}
\caption{$P_{D_0}$ vs. $P_{F_0}$ for all the presented set of MRC (MRC, mMRC, TR-MRC, TR-mMRC) and WL (WL, TR-WL) rules with $(N, K) = (32, 8)$ with mean of shadowing effect $\mu_{\lambda} = 6$ dB (indoor-to-outdoor) and a path-loss exponent of 2 in the network.}
\label{FIG3b}
\end{figure} 

\begin{figure}[t]
\vspace*{-5mm}
\begin{center}
 \includegraphics[width=1.9\linewidth]{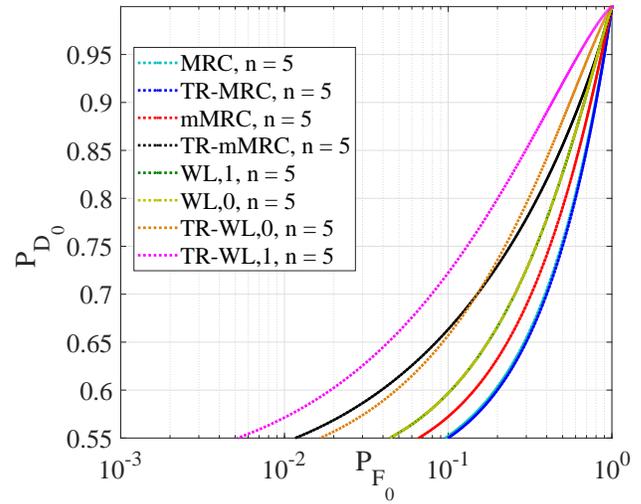}
\end{center}
\vspace*{-6mm}
\caption{$P_{D_0}$ vs. $P_{F_0}$ for all the presented set of MRC (MRC, mMRC, TR-MRC, TR-mMRC) and WL (WL, TR-WL) rules with $(N, K) = (64, 8)$ with a path-loss exponent $n = 5$ (indoor/outdoor) and $\mu_{\lambda} = 4$ dB.}
\label{FIG4a}
\end{figure} 

\begin{figure}[t]
\vspace*{-5mm}
\begin{center}
 \includegraphics[width=1.9\linewidth]{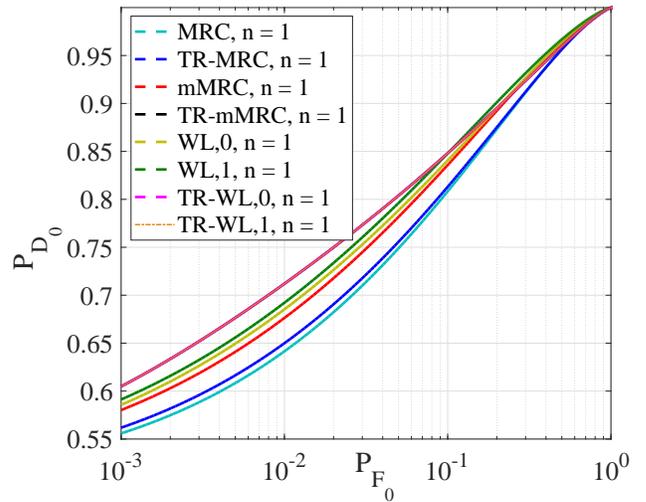}
\end{center}
\vspace*{-6mm}
\caption{$P_{D_0}$ vs. $P_{F_0}$ for all the presented set of MRC (MRC, mMRC, TR-MRC, TR-mMRC) and WL (WL, TR-WL) rules with $(N, K) = (64, 8)$ with path-loss exponent $n = 1$ (tunnel-like environment) and $\mu_{\lambda} = 4$ dB.}
\label{FIG4b}
\end{figure} 

Fig.~\ref{FIG3a}, Fig.~\ref{FIG3b}, Fig.~\ref{FIG4a} and Fig.~\ref{FIG4b} exhibit the large scale channel effects between the SUs and DFC for all the presented set of rules. As the mean of the shadowing distribution $\mu_{\lambda}$ increases, the channel encounters a very small group of scattering clusters resulting in a decrease in the mean signal level attenuation. It is evident from Fig.~\ref{FIG3a}, for a low $\mu_{\lambda} = 2$ dB (indoor environment congested with different groups of scatterers), TR-MRC and TR-mMRC versions are unable to combat the severe shadowing effects, but can offer some improvement over MRC for $\mu_{\lambda} \geq 4$ dB (open indoor environment and outdoor-to-indoor case or vice versa). However, TR-WL do offer some improvement in performance even in severe shadowing condition. TR-WL also do come out a winner in environments with high propagation path-loss (high $n = 5$) over TR-mMRC technique (refer to Fig.~\ref{FIG4a}). It is to be noted here that for the results generated in Fig.~\ref{FIG3a} and Fig.~\ref{FIG3b}, we consider a moderate number of SUs, $K = 8$ and moderate size array at the DFC ($N = 32$), while for Fig.~\ref{FIG4a} and Fig.~\ref{FIG4b} large size array is considered ($N = 64$) at the DFC.

\begin{figure}[t]
\hspace*{-150mm}
\vspace*{-4mm}
\begin{center}
 \includegraphics[width=1.05\linewidth]{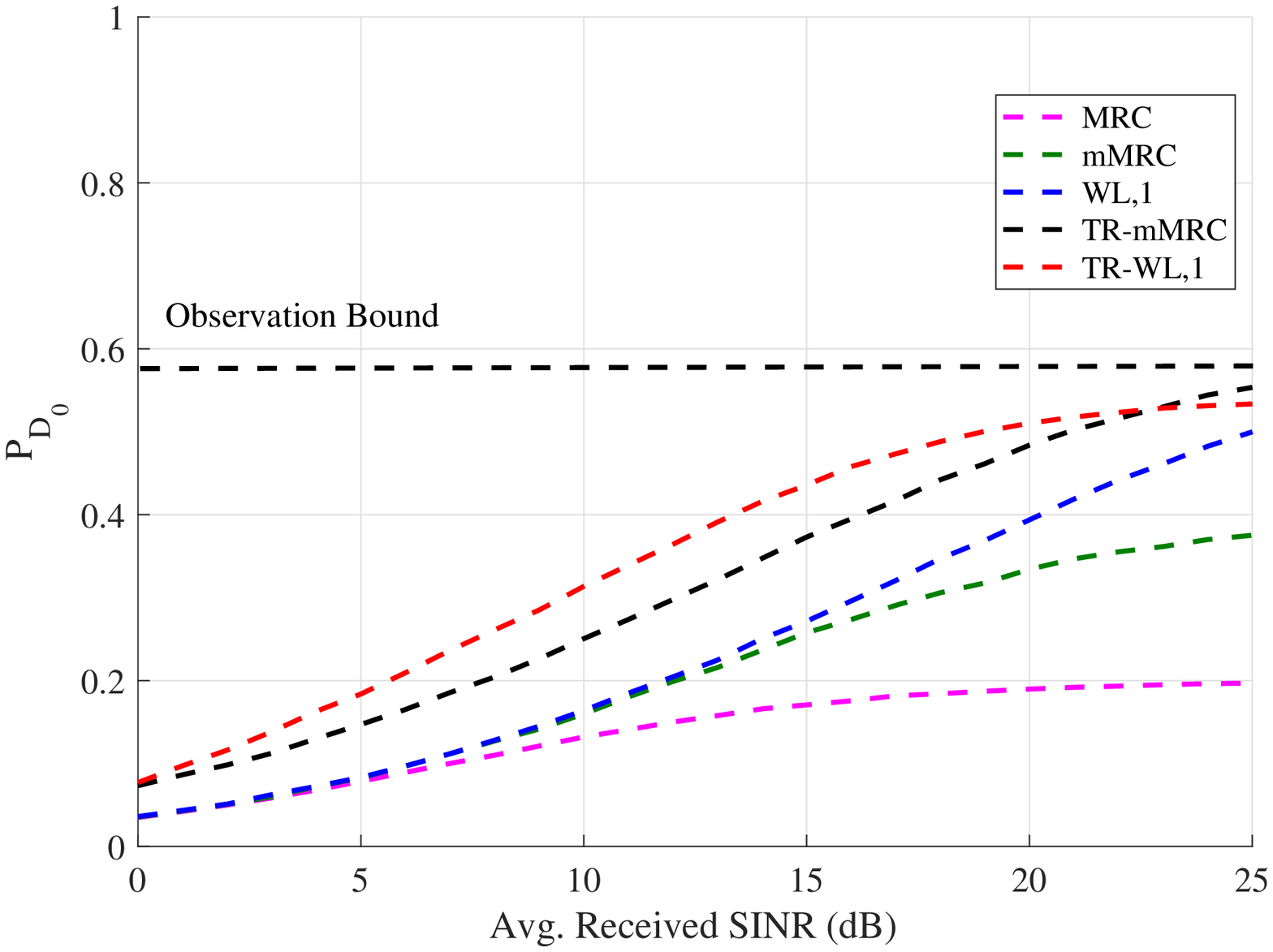}
\end{center}
\vspace*{-6mm}
\caption{$P_{D_0}$ vs. channel SINR (dB) for few of the presented rules with $(N, K) = (32, 8)$ over $(\mu_{\lambda}, \sigma_{\lambda}) = (4, 2)$ dB and a path-loss exponent of 2; $P_{F_0} \leq 0.01$ and $(P_{D, k}, P_{F, k}) = (0.5, 0.05)$.}
\label{FIG9}
\vspace*{-4mm}
\end{figure} 

\begin{figure}[t]
\hspace*{-100mm}
\begin{center}
 \includegraphics[width=1.09\linewidth]{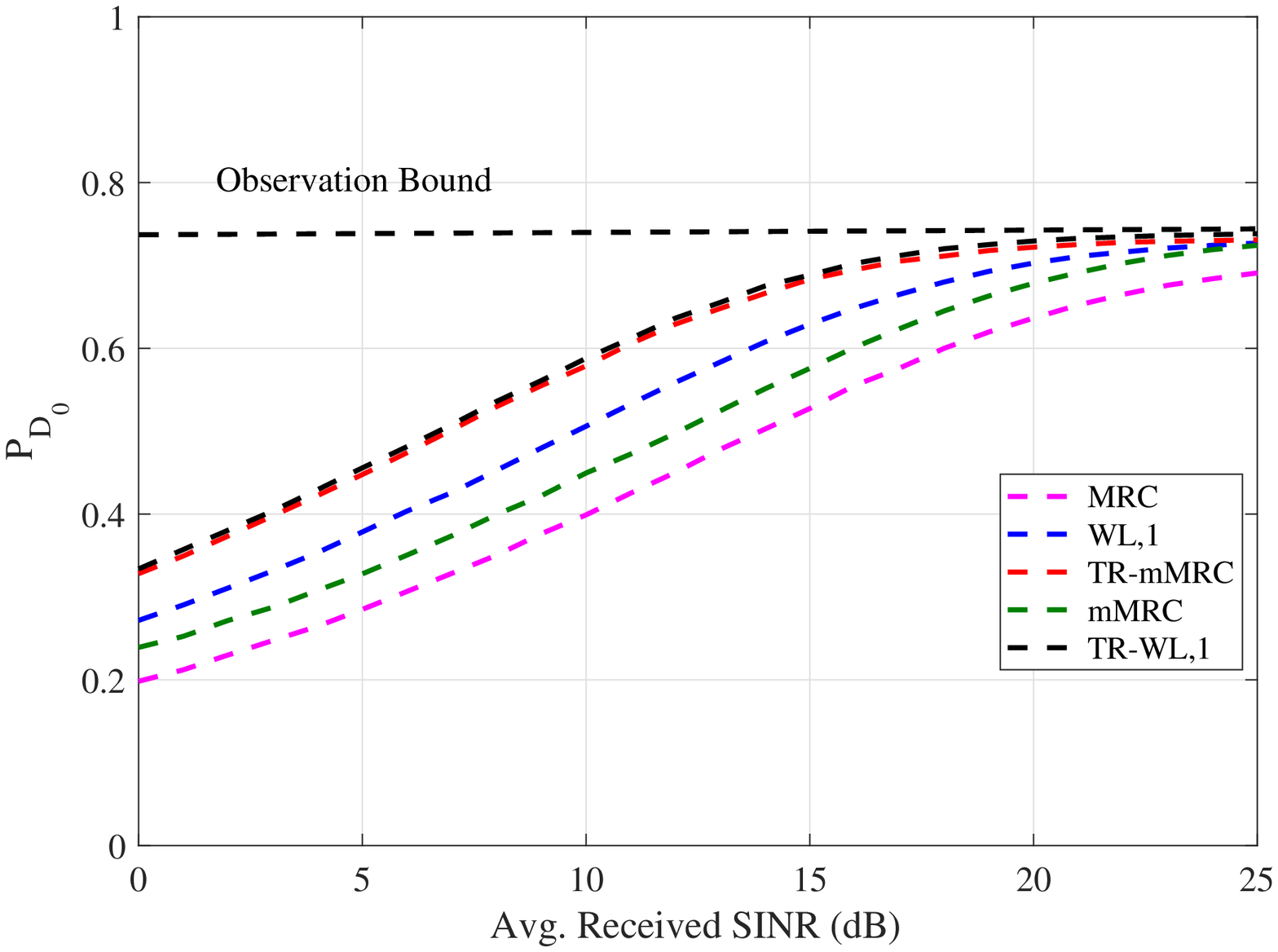}
\end{center}
\vspace*{-6mm}
\caption{$P_{D_0}$ vs. channel SINR (dB) for few of the presented rules with $(N, K) = (64, 8)$ over $(\mu_{\lambda}, \sigma_{\lambda}) = (4, 2)$ dB and a path-loss exponent of 2; $P_{F_0} \leq 0.01$ and $(P_{D, k}, P_{F, k}) = (0.5, 0.05)$.}
\label{FIG10}
\end{figure} 

\begin{figure}[t]
\hspace*{-100mm}
\vspace*{-4mm}
\begin{center}
 \includegraphics[width=1.09\linewidth]{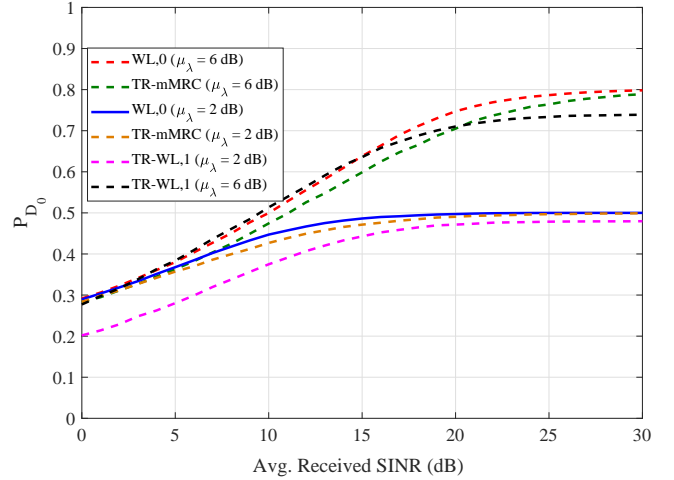}
\end{center}
\vspace*{-8mm}
\caption{$P_{D_0}$ vs. channel SINR (dB) for few of the presented rules with $(N, K) = (64, 8)$ where the curves are generated by varying path-loss exponent $n$ (tunnel-like environment $n = 1$, indoor/outdoor $n = 5$) with $\mu_{\lambda} = 4$ dB; $P_{F_0} \leq 0.01$ and $(P_{D, k}, P_{F, k}) = (0.5, 0.05)$.}
\label{FIG11}
\vspace*{-30mm}
\end{figure}

\begin{figure}[t]
\vspace*{-4mm}
\hspace*{-100mm}
\begin{center}
 \includegraphics[width=1.09\linewidth]{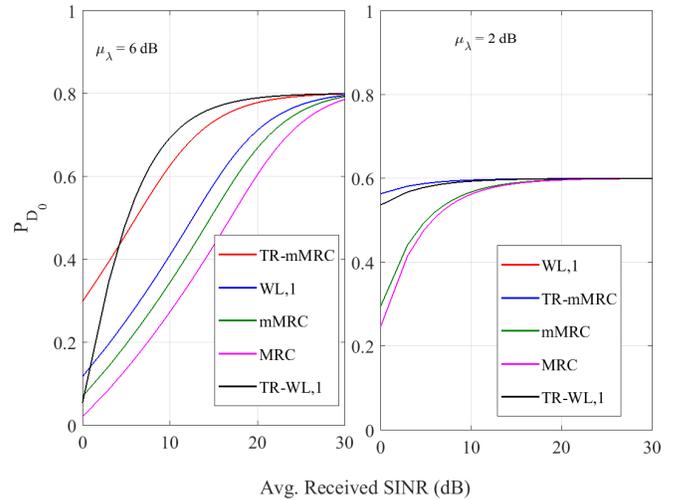}
\end{center}
\vspace*{-6mm}
\caption{$P_{D_0}$ vs. channel SINR (dB) for few of the presented rules with $(N, K) = (64, 8)$ where the curves are generated by varying mean of shadowing effect $\mu_{\lambda}$ (indoor $\mu_{\lambda} = 2$ dB, indoor-to-outdoor $\mu_{\lambda} = 6$ dB) with a path-loss exponent of 2 in the network; $P_{F_0} \leq 0.01$ and $(P_{D, k}, P_{F, k}) = (0.5, 0.05)$.}
\label{FIG12}
\vspace*{-4mm}
\end{figure}

In Figs.~\ref{FIG9}, \ref{FIG10}, \ref{FIG11} and \ref{FIG12}, we plot for few of the presented rules, $P_{D_0}$ as a function of the channel SINR (dB) under $P_{F_0} \leq 0.01$ under wideband cooperative spectrum sensing among $K = 8$ SUs. We investigate the effect of multiple receive antennas at the DFC, with moderately high $N = 32$ (Fig.~\ref{FIG9}) and very large $N = 64$ (Fig.~\ref{FIG10}). Firstly, these numerical results confirm $P_{D_0}$ v/s $P_{F_0}$ performances of few of the presented set of rules, i.e., TR-mMRC and TR-WL,1 outperforms any other form of MRC and WL rules, with TR-WL,1 being the winner. When the DFC employs moderately high $N = 32$ number of receive antennas, MRC and mMRC rules never approach the observaton bound even not in case of high SINR. But as $N$ increases to 64, MRC and mMRC also approach close to the observation bound for higher SINRs. It is also evident from Fig.~\ref{FIG11}, that in presence of high $N$, the TR variants of the fusion rules do not offer considerable advantage over other fusion rules in presence of high propagation pathloss ($n = 5$). For high shadowing mean $\mu_{\lambda} = 6$ dB, TR fusion rules outperform any other DF rules by a large margin. But in presence of large number of scatterers with high shadowing effect, WL rules perform very close to the TR set (refer to Fig.~\ref{FIG12}). Hence, we can broadly conclude that TR based fusion rules are efficient for spectrum sharing among interfering SUs in presence of fading channels, but do not perform that well in presence of severe shadowing.

\section{Conclusions}\label{S6}

In this paper, we have considered OFDM-based wideband collaborative spectrum sensing using a DFC equipped with multiple receive antennas. 
We have eliminated the use of CP from our consideration in order to improve overall spectrum efficiency.
We have proposed TR-MRC, TR-mMRC and TR-WL rules for decision fusion in order to mitigate wideband channel effects like frequency-selective fading and inter-carrier interference. 
Simulation results demonstrate considerable enhancement in performance is in case of TR based fusion rules over conventional ones. 
The TR versions also perform better than WL rules in environments suffering from high propagation path-loss but are unable to mitigate severe shadowing effects considerably; only a slight improvement is offered by TR-WL.
In future, we will analyze how these proposed massive MIMO based DF rules perform in presence of correlated observations and reporting over the frequency bins, and investigate the benefit of exploiting multi-slot decisions to capitalize the benefits of time-integration in sensing performance, following \cite{chawla2019}.
Furthermore, we intend to conduct a first-of-a-kind indoor measurement campaign to capture the propagation characteristics in a `virtual' massive MIMO based radio networks and compare the performance of the proposed DF rules over the measured channel. 

\appendices
\renewcommand{\theequation}{\thesection.\arabic{equation}}

\section{Proof of Propositions 1 and 2}

We first start with the argument that since $P(\mathbf{y}^l|\mathbf{G}^l, \mathcal{H}_i^l)$ is assumed to follow Gaussian mixture distribution, $\mathbf{\Gamma}^{l, \text{WB}}_{i, \text{rule}}|\mathbf{G}^l, \mathcal{H}_i^l$ is also distributed according to Gaussian mixture model that is,
\begin{align} \label{eq37}
&\mathbf{\Gamma}^{l, \text{WB}}_{i, \text{rule}}|\mathbf{G}^l, \mathcal{H}_i^l~\sim\sum_{\mathbf{x}^l \in {\mathbf{\chi}}^K} P({x}^l|\mathcal{H}_i^l) \nonumber\\
&~~~~~~~~\mathcal{N}\big(\mathbb{E}\big\{\mathbf{\Gamma}^{l, \text{WB}}_{i, \text{rule}}|\mathbf{G}^l, \mathbf{x}^l\big\}, \mathbb{V}\big\{\mathbf{\Gamma}^{l, \text{WB}}_{i, \text{rule}}|\mathbf{G}^l, \mathbf{x}^l\big\}\big)
\end{align}
Using Gaussian moment matching \cite{barshalom2004}, we can approximate the pdf in (\ref{eq37}) as,
\begin{align} \label{eq38}
&\mathbf{\Gamma}^{l, \text{WB}}_{i, \text{rule}}|\mathbf{G}^l, \mathcal{H}_i^l~\overset{\text{approx}}{\sim}\mathcal{N}\big(\mathbb{E}\big\{\mathbf{\Gamma}^{l, \text{WB}}_{i, \text{rule}}|\mathbf{G}^l, \mathcal{H}_i^l\big\}, \mathbb{V}\big\{\mathbf{\Gamma}^{l, \text{WB}}_{i, \text{rule}}|\mathbf{G}^l, \mathcal{H}_i^l\big\}\big).
\end{align}
Since at low-SNR, the components of the Gaussian mixture gets concentrated within a certain region. To prove Propositions 1 and 2, we need to evaluate the mean and variance of $\mathbf{\Gamma}^{l, \text{WB}}_{i, \text{rule}}|\mathbf{G}^l, \mathcal{H}_i^l$ separately for the WL and MRC rules. For this purpose, let us define, $\tilde{\mathbf{G}}^l \overset{\triangle}{=} \big[{{\mathbf{G}}^l}^t~~{{\mathbf{G}}^l}^{\dagger}\big]^t$ and ${\tilde{\mathbf{a}}^l_{\text{m}}} \overset{\triangle}{=} \frac{1}{2}\big[{\mathbf{a}^l_{\text{m}}}^t~~{\mathbf{a}^l_{\text{m}}}^{\dagger}\big]^t$, where $\mathbf{a}^l_{\text{m}}$ is either $\mathbf{a}^l_{\text{MRC}}$, $\mathbf{a}^l_{\text{mMRC}}$ or $\mathbf{a}^l_{\text{TR-mMRC}}$ depending on the fusion rules chosen from the set of MRC and TR-MRC rules.

\subsection{Proof of Proposition 1}

First of all, we evaluate the mean and variance of $\mathbf{\Gamma}^{l, \text{WB}}_{i, \text{WL}}|\mathbf{G}^l, \mathcal{H}_i^l$ as,
\begin{align} \label{eq39}
&\mathbb{E}\big\{\mathbf{\Gamma}^{l, \text{WB}}_{i, \text{WL}}|\mathbf{G}^l, \mathbf{x}^l\big\} = \sum_{\mathbf{x}^l \in {\mathbf{\chi}}^K} P({x}^l|\mathcal{H}_i^l) \mathbb{E}\big\{\mathbf{\Gamma}^{l, \text{WB}}_{i, \text{WL}}|\mathbf{G}^l, \mathbf{x}^l\big\} \nonumber\\
&~~~= \sum_{\mathbf{x}^l \in {\mathbf{\chi}}^K} P({x}^l|\mathcal{H}_i^l)\frac{\sqrt{\rho^l} (\mathbf{\mu}_i^l)^t (\mathbf{G}^l)^{\dagger} \mathbf{\Sigma}^{- 1}_{\mathbf{y}^l|\mathbf{G}^l, \mathcal{H}^l_i} \mathbf{G}^l \mathbf{x}^l}{||\mathbf{\Sigma}^{- 1}_{\mathbf{y}^l|\mathbf{G}^l, \mathcal{H}^l_i} \mathbf{G}^l \mathbf{x}^l||} \nonumber\\
&~~~= \frac{\sqrt{\rho^l} (\mathbf{\mu}_i^l)^t (\mathbf{G}^l)^{\dagger} \mathbf{\Sigma}^{- 1}_{\mathbf{y}^l|\mathbf{G}^l, \mathcal{H}^l_i} \mathbf{G}^l~\mathbb{E}\{\mathbf{x}^l|\mathcal{H}_i^l\}}{||\mathbf{\Sigma}^{- 1}_{\mathbf{y}^l|\mathbf{G}^l, \mathcal{H}^l_i} \mathbf{G}^l \mathbf{x}^l||}
\end{align}
and
\begin{align} \label{eq40}
&\mathbb{V}\big\{\mathbf{\Gamma}^{l, \text{WB}}_{i, \text{WL}}|\mathbf{G}^l, \mathbf{x}^l\big\} \nonumber\\
&~~~= \sum_{\mathbf{x}^l \in {\mathbf{\chi}}^K} P({x}^l|\mathcal{H}_i^l) \mathbb{E}\big\{||\mathbf{\Gamma}^{l, \text{WB}}_{i, \text{WL}} - {\tilde{\mathbf{G}}}^l~\mathbb{E}\{\mathbf{x}^l|\mathcal{H}_i^l\}||^2|\mathbf{G}^l, \mathbf{x}^l\big\} \nonumber\\
&~~~= \sum_{\mathbf{x}^l \in {\mathbf{\chi}}^K}{\tilde{\mathbf{G}}}^l~\mathbb{E}\{(\mathbf{x}^l - \mathbb{E}\{\mathbf{x}^l|\mathcal{H}_i^l\}) \nonumber\\
&~~~~~~~~~~\times (\mathbf{x}^l - \mathbb{E}\{\mathbf{x}^l|\mathcal{H}_i^l\})^T|\mathcal{H}_i^l\} ({\tilde{\mathbf{G}}}^l)^{\dagger} + 2 \sigma^2_{e,l}
\end{align}
Under simplifying assumptions of $\mathbb{E}\{\mathbf{x}^l|\mathcal{H}_0^l\} = (2 P_{F}^{l} - 1) \mathbf{1}_K$ (\ref{eq40}) becomes,
\begin{align} \label{eq41}
\mathbb{V}\big\{\mathbf{\Gamma}^{l, \text{WB}}_{i, \text{WL}}|\mathbf{G}^l, \mathbf{x}^l\big\} &= [1 - (2 P_{F}^{l} - 1)^2]{\tilde{\mathbf{G}}}^l ({\tilde{\mathbf{G}}}^l)^{\dagger} + 2 \sigma^2_{e,l} \nonumber\\
&\approx \lim_{K \to \infty} 2(1 - {\delta^l}^2)K + 2 \sigma^2_{e,l}
\end{align}
where $\delta^l = (2 P_{F}^{l} - 1)$. Using (\ref{eq39}) and (\ref{eq41}) and exploiting (\ref{eq38}), we obtain the low-SINR approximation for $P_{F_0}^l$ as,
\begin{align} \label{eq42}
P_{F_0}^l &\approx Q \Bigg(\frac{\gamma^l - \frac{\sqrt{\rho^l} (\mathbf{\mu}_i^l)^t (\mathbf{G}^l)^{\dagger} \mathbf{\Sigma}^{- 1}_{\mathbf{y}^l|\mathbf{G}^l, \mathcal{H}^l_i} \mathbf{G}^l~\mathbb{E}\{\mathbf{x}^l|\mathcal{H}_i^l\}}{||\mathbf{\Sigma}^{- 1}_{\mathbf{y}^l|\mathbf{G}^l, \mathcal{H}^l_i} \mathbf{G}^l \mathbf{x}^l||}}{\sqrt{2(1 - {\delta^l}^2)K + 2 \sigma^2_{e,l}}}\Bigg) \nonumber\\
&\approx \lim_{N \to \infty} Q \Bigg(\frac{\gamma^l - \frac{N \delta^l\sqrt{2\rho^l} (\mathbf{\mu}_i^l)^t \mathbf{x}^l \mathbf{V}_i^l \mathbf{D}_g^l}{\sigma_{e, l}\sqrt{(\mathbf{\mu}_i^l)^t \mathbf{V}_i^l \mathbf{D}_g^l (\mathbf{V}_i^l)^t \mathbf{\mu}_i^l}}}{\sqrt{2(1 - {\delta^l}^2)K + 2 \sigma^2_{e,l}}}\Bigg)
\end{align}
Under simplifying assumptions of $\mathbb{E}\{\mathbf{x}^l|\mathcal{H}_0^l\} = (2 P_{F}^{l} - 1) \mathbf{1}_K \overset{\triangle}{=} \delta^l\mathbf{1}_K$ and $Z = K$, (\ref{eq42}) simplifies to,
\begin{align} \label{eq43}
P_{F_0}^l &\approx Q \Bigg(\frac{\gamma^l - \frac{2N \delta^l \sqrt{\rho^l} \sum_{k =1}^K \lambda_k \beta_k^l(k -1) \mu_{1, 0, k, l}}{\sqrt{\sum_{k =1}^K \lambda_k \beta_k^l(k -1) } \mu^2_{1, 0, k, l}}}{\sqrt{2(1 - {\delta^l}^2)K + 2 \sigma^2_{e,l}}}\Bigg)
\end{align}
which can be easily inverted to (\ref{eq19}).

\subsection{Proof of Proposition 2}

First of all, we evaluate the mean and variance of $\mathbf{\Gamma}^{l, \text{WB}}_{i, \text{m}}|\mathbf{G}^l, \mathcal{H}_i^l$ as,
\begin{align} \label{eq44}
\mathbb{E}\big\{\mathbf{\Gamma}^{l, \text{WB}}_{i, \text{m}}|\mathbf{G}^l, \mathbf{x}^l\big\} &= \sum_{\mathbf{x}^l \in {\mathbf{\chi}}^K} P({x}^l|\mathcal{H}_i^l) \mathbb{E}\big\{\mathbf{\Gamma}^{l, \text{WB}}_{i, \text{m}}|\mathbf{G}^l, \mathbf{x}^l\big\} \nonumber\\
&= \sum_{\mathbf{x}^l \in {\mathbf{\chi}}^K} P({x}^l|\mathcal{H}_i^l) \sqrt{\rho^l}\mathbb{R}\big\{(\mathbf{a}^l_{\text{m}})^{\dagger} \mathbf{G}^l \mathbf{x}^l\big\} \nonumber\\
&= \sqrt{\rho^l}\mathbb{R}\big\{(\mathbf{a}^l_{\text{m}})^{\dagger} \mathbf{G}^l ~\mathbb{E}\{\mathbf{x}^l|\mathcal{H}_i^l\}\big\}
\end{align}
and
\begin{align} \label{eq45}
&\mathbb{V}\big\{\mathbf{\Gamma}^{l, \text{WB}}_{i, \text{m}}|\mathbf{G}^l, \mathbf{x}^l\big\} \nonumber\\
&~~~= \sum_{\mathbf{x}^l \in {\mathbf{\chi}}^K} P({x}^l|\mathcal{H}_i^l) \mathbb{E}\big\{||\mathbf{\Gamma}^{l, \text{WB}}_{i, \text{m}} - {\tilde{\mathbf{G}}}^l~\mathbb{E}\{\mathbf{x}^l|\mathcal{H}_i^l\}||^2|\mathbf{G}^l, \mathbf{x}^l\big\} \nonumber\\
&~~~= \sum_{\mathbf{x}^l \in {\mathbf{\chi}}^K} (\tilde{\mathbf{a}}^l_{\text{m}})^{\dagger}{\tilde{\mathbf{G}}}^l~\mathbb{E}\{(\mathbf{x}^l - \mathbb{E}\{\mathbf{x}^l|\mathcal{H}_i^l\}) \nonumber\\
&~~~~~~~~~~\times (\mathbf{x}^l - \mathbb{E}\{\mathbf{x}^l|\mathcal{H}_i^l\})^T|\mathcal{H}_i^l\} ({\tilde{\mathbf{G}}}^l)^{\dagger}\tilde{\mathbf{a}}^l_{\text{m}} + \frac{\sigma^2_{e,l}}{2}||\mathbf{a}^l_{\text{m}}||^2
\end{align}
Under simplifying assumptions of $\mathbb{E}\{\mathbf{x}^l|\mathcal{H}_0^l\} = (2 P_{F}^{l} - 1) \mathbf{1}_K$ (\ref{eq45}) becomes,
\begin{align} \label{eq46}
\mathbb{V}\big\{\mathbf{\Gamma}^{l, \text{WB}}_{i, \text{m}}|\mathbf{G}^l, \mathbf{x}^l\big\} 
&\approx \lim_{K \to \infty} \sqrt{1/2((1 - {\delta^l}^2)K + \sigma^2_{e,l})}~||\mathbf{a}^l_{\text{m}}||
\end{align}
where $\delta^l = (2 P_{F}^{l} - 1)$. Using (\ref{eq44}) and (\ref{eq46}) and exploiting (\ref{eq38}), we obtain the low-SINR approximation for $P_{F_0}^l$ as,
\begin{align} \label{eq47}
P_{F_0}^l &\approx Q \Bigg(\frac{\gamma^l - \sqrt{\rho^l}\mathbb{R}\big\{(\mathbf{a}^l_{\text{m}})^{\dagger} \mathbf{G}^l ~\mathbb{E}\{\mathbf{x}^l|\mathcal{H}_i^l\}\big\}}{\sqrt{1/2((1 - {\delta^l}^2)K + \sigma^2_{e,l})}~||\mathbf{a}^l_{\text{m}}||}\Bigg) \nonumber\\
&\approx \lim_{N \to \infty} Q \Bigg(\frac{\gamma^l - \sqrt{N\rho^l}\delta^l ||\mathbf{a}^l_{\text{m}}||^2}{\sqrt{1/2((1 - {\delta^l}^2)K + \sigma^2_{e,l})}~||\mathbf{a}^l_{\text{m}}||}\Bigg)
\end{align}
Under simplifying assumptions of $\mathbb{E}\{\mathbf{x}^l|\mathcal{H}_0^l\} = (2 P_{F}^{l} - 1) \mathbf{1}_K \overset{\triangle}{=} \delta^l\mathbf{1}_K$ and $Z = K$, (\ref{eq47}) simplifies to (\ref{eq35}) for each set of MRC and TR-MRC rules.

\bibliographystyle{IEEEtran}
\bibliography{bibliography}

\vspace*{-5mm}

\end{document}